\documentclass[aps, prfluids, 11pt, superscriptaddress,amsfonts,amssymb,amsmath]{revtex4-2}
\usepackage{times}
\usepackage{graphicx,xcolor}
\usepackage{epstopdf,epsfig}
\newcommand{\mi}{\mathrm{i}}
\newcommand*\diff{\mathop{}\!\mathrm{d}}
\usepackage[normalem]{ulem}
\usepackage{bm}
\usepackage{relsize}
\allowdisplaybreaks
\usepackage[percent]{overpic}
\DeclareRobustCommand{\rchi}{{\mathpalette\irchi\relax}}
\newcommand{\irchi}[2]{\raisebox{\depth}{$#1\chi$}} 

\newcommand{\Rey}{{\rm Re}}

\makeatletter
\def\blfootnote{\xdef\@thefnmark{}\@footnotetext}
\makeatother

\begin{document}


\title{\Large Perturbation symmetries in shear-thinning viscoelastic pipe flows and the Petrov--Galerkin implementation}


\author{M. Malik}
\email[]{dr.malik.barak@gmail.com}
\affiliation{Singapore University of Technology and Design, 8 Somapah Road, Singapore 487372}
\altaffiliation{}
\author{Martin Skote}
\email[]{m.skote@cranfield.ac.uk}
\affiliation{Cranfield University, College Road, Cranfield MK43 0AL, United Kingdom}
\author{Roland Bouffanais}
\email[]{bouffanais@sutd.edu.sg}
\affiliation{Singapore University of Technology and Design, 8 Somapah Road, Singapore 487372}


\begin{abstract}
The perturbations of the laminar shear-thinning viscoelastic pipe flow under Finitely Extensible Nonlinear Elastic model with Peterlin
approximation (FENE-P) are shown to exhibit leading-order power-law behaviours, and the expected odd-even parities with respect to the radial coordinate that depend on the azimuthal wave\-number, $n$. The analysis helps regularizing the governing system of equations at the centreline, and allows for a complete stability analysis of three-dimensional perturbations for a general integer value of $n$, which has hitherto remained a challenge for FENE-P models. It is shown here that the symmetry and analytic behaviours of the velocity and pressure fields of the Newtonian counterpart are both preserved in this flow, and the reason is elucidated. For  $|n|=1$, the perturbations to the correlations between the axial component and the radial or azimuthal components of the end-to-end polymer vector exhibit behaviour similar to that of the velocity perturbations close to the centreline, and are traced to the uniformity of axial traction with respect to the azimuthal direction. For all values of $n$, the fluctuation to the end-to-end length of the polymer chain vanishes at the centreline. The ansatzes for the perturbations to the components of conformation tensor arrived here, using heuristics, are later proved in two separate ways: Frobenius method, and a method that utilizes observations from Fourier analysis. We also reveal the existence of natural modes of polymers with trivial velocity perturbations in the limit of vanishing polymer viscosity. The complex frequency spectrum of these modes is continuous with radial stratification. Finally, the ansatzes are implemented using a Petrov--Galerkin spectral scheme. 
\end{abstract}
\keywords{Viscoelasticity \and Instabilities \and Petrov--Galerkin \and Regularization \and Polar singularity}
\maketitle

\section{Introduction}\label{sec:introduction}
Over the past decades, the well-known phenomenon of turbulent drag reduction associated with the addition of heavy-molecular long-chain polymers~\cite{white2008mechanics,graham2014drag} has motivated research on the emergence of modal instabilities. Understanding the onset of instabilities could enhance the prediction of transition in such shear-thinning fluids~\cite{guzel2009predicting}. The modal analysis of disturbances remains important as it sheds light on the underlying physical phenomenon in a mathematically tractable manner both in the case of infinitesimal perturbations, or when coming to analysing the flow field obtained from CFD or experiments (see for example, Refs.~\cite{theofilis2011global,taira2017modal}). 
Recently, the elasto-inertial turbulence (EIT)~\cite{samanta2013elasto,dubief2013mechanism,chandra2018onset,dubief2020first} has been identified to be connected to modal instabilities~\cite{page2020exact,shekar2020self}. This EIT has been postulated~\cite{xi2012intermittent} and confirmed~\cite{choueiri2018exceeding} to be the phenomenon behind the maximum-drag-reduction asymptote commonly observed in experiments.

The disturbances in shear-thinning fluids can be analysed either by using generalized Newtonian fluid (GNF) models, such as the Carreau model~\cite{chikkadi2005preventing,nouar2007delaying}, or by considering a range of dumbbell constitutive models, such as the Oldroyd-B, or the more elaborate FENE-CR and FENE-P models. Given the theoretical backing that such dumbbell models receive from the kinetic theory~\cite{phan2017understanding}, we focus our attention on one of these canonical models, namely the FENE-P (which stands for ``Finitely Extensible Nonlinear Elastic model with Peterlin approximation") model for this paper as it predicts shear thinning. Further, when the GNF models are to be used, substantial care is needed to define the Reynolds number that takes into account of the varying effective viscosity~\cite{guzel2009predicting}.

The vast majority of linear stability analyses that use dumbbell models were based on geometries with Cartesian symmetries~\cite{sureshkumar1995linear,hoda2008energy,zhang2013linear,agarwal2014linear,ray2015absolute,wilson2015linear,castillo2017towards}. Among those performed in cylindrical coordinates, the analyses systematically excluded the point of singularity at the null radial coordinate~\cite{larson1990purely}. Although the pipe geometry finds a commonplace in applications, it has been largely excluded from non-Newtonian studies due to the difficulty of specifying the regularity conditions at the pipe centre. 

On the other hand, such regularity conditions are well known in the case of Newtonian pipe flows. In one way of treating this singular behaviour, the conditions of uniformity of velocity fields along the azimuthal direction in the limit of reaching the centre~\cite{schmid2001stability,khorrami1989application} are used as boundary conditions. Another approach consists of circumventing this singular point by using the symmetries of the perturbations predicted by its analytic behaviour while the domain is artificially extended to $-1\leq r/R \leq 1$, where $R$ is the pipe radius~\cite{priymak1998accurate,meseguer2000spectral,meseguer2003linearized}. A third way for treating this singularity, still in the Newtonian case, consists of deriving the regularity conditions at the pipe centre ($r=0$) by making use of the perturbations analytic behaviour~\cite{malik2019linear}. For an early implementation of such requirement from analyticity, see Duck~\cite{duck1983flow}.

In most stability analyses of non-Newtonian flows performed in circular geometry \cite{chen1991interfacial,chen1992elastic,chen1993stability,ye2016instability,garg2018viscoelastic}, only axisymmetric disturbances with the azimuthal wavenumber $n = 0$ were studied under upper-convected Maxwell (UCM) or Oldroyd-B models. Studying this particular mode does not pose a problem, since the terms exhibiting an explicit dependence in $nr^{-1}$ in the linearized governing equations for conformation tensor would vanish. 

Miller and Rallison~\cite{miller2007instability} consider the mode with $n=1$ besides $n=0$ for the UCM and Oldroyd-B models. For $n=1$, the azimuthal perturbation velocity, $w'$, and radial perturbation velocity, $v'$, follow a relation $v'+iw' = 0$ at the centreline, a result observed by Khorrami {\it et al.}~\cite{khorrami1989application} and established by Lewis and Bellan~\cite{lewis1990physical} as a property of any vector in all physical problems in the plane perpendicular to the polar axis. This fact, together with the property that $v'(r)$ and $w'(r)$ are even, sets $nr^{-1}(v'+iw') =0$ in the governing equations for polymer stress components. 

However, carrying out a generalized stability analysis with $|n|\geq 2$ will warrant more robust regularity conditions. For modes with $|n|\geq 2$, this singularity could still be removable if the leading order power-law behaviour with respect to radial coordinate is known. When the power-law behaviour of the unknowns would remain unaccounted explicitly in the analysis, the terms with the factors of $nr^{-1}u'$, $nr^{-1}v'$ or $nr^{-1}w'$ (where $u'$ is the axial perturbation velocity) in the governing equations would pose difficulty even when the vanishing conditions, namely, $u'=v'=w'=0$ at $r=0$ found by the analysis of Khorrami {\it et al.}~\cite{khorrami1989application} are used. 

Further, neither the UCM and nor the Oldroyd-B models considered in Miller and Rallison~\cite{miller2007instability} account for shear-thinning, hence the effect of shear-rate dependent viscosity would be obscured under such models. 

More recently, Malik {\it et al.}~\cite{malik2020viscoelastic} observed that the laminar profiles of the velocity field and conformation tensor of the FENE-P viscoelastic pipe flow exhibit odd or even symmetries. Moreover, Malik {\it et al.}~\cite{malik2020viscoelastic} postulated that these properties of the laminar profiles can be exploited to derive the symmetries and leading-order power-law behaviours of the perturbations to this flow similarly to the way it is classically achieved in the Newtonian case. These properties can eventually be exploited to circumvent the singularity at the pipe centre as carried out by Priymak and Miyazaki~\cite{priymak1998accurate} and Meseguer and Trefethen~\cite{meseguer2003linearized}, or alternatively, to derive a set of more robust regularity conditions for all values of $n$ as in Malik {\it et al.}~\cite{malik2019linear}. 

In the present paper, we begin by presenting the mean-flow variables of the flow under FENE-P model in Sect.~\ref{sec:model}, and proceed to Sect.~\ref{sec:linsys} where we derive the leading-order power-law behaviour, and analyse the symmetries of the perturbations under the FENE-P constitutive viscoelastic model. We find that the perturbations indeed exhibit odd/even parities depending on the value of $n$. In fact, the change in parities with $n$ is to be expected in any physical systems. Lewis and Bellan~\cite{lewis1990physical} noted by the analysis of Fourier modes that the parities of scalars and vectors in axial planes can be arrived without the need to analyse the governing equations. However, the complexity gets enhanced when these physical quantities of vectors in nature (velocity, for example) forces certain tensors. As we show here by Frobenius method, the leading order behaviour of the tensor components could be decided by the physical problem (i.e., governing equations) unlike the vectors in axial planes and scalars.

{
In Sect.~\ref{sec:valid}, we give a numerical validation for the power-law behaviour of the perturbations to the conformation tensor for all cases of $(n, \alpha)$ in the limit of vanishing polymer viscosity using a scheme that does not use the ansatzes intended to be verified.
}

We note in Sect.~\ref{sec:signific} that the perturbations to the correlations between the axial component and the radial or azimuthal components of the polymer end-to-end vector exhibit the same behaviour as the perturbation velocities in these respective directions close to centreline, which become evident for $|n|=1$ as shown in Sect.~\ref{sec:linsys}. Since the Oldroyd-B model can be seen as a special case of the FENE-P model---with the Peterlin function set equal to unity---these derived properties can be more generally applied to various dumbbell models~\cite{phan2017understanding}.

{
In Sect.~\ref{sec:natural}, we show that there are {\it natural} modes of polymers with non-trivial perturbations to the components of the conformation tensor that have trivial velocity perturbations for the case $\beta = 1$ and that these cannot exist for $\beta \neq 1$. We also derive an analytical expression for the corresponding spectrum of complex frequencies that shows that this spectrum is continuous in nature and are stratified in the radial direction.

In Sect.~\ref{sec:petrov}, we describe a Petrov--Galerkin scheme for implementation of the derived ansatzes and explain the advantage of this method over other available methods for the FENE-P model. We also specify trial-basis functions that follow the derived ansatzes, and the test functions that are needed to project the governing system of equations on. We find that the instability occurs as long as the Weissenberg number is not too low. The spectrum of complex frequencies has branches in the complex plane that exhibits similarity to the natural spectrum described in Sect.~\ref{sec:natural}. 
}

\section{FENE-P model and laminar profiles}\label{sec:model}
The unsteady flow of dilute polymers modelled as dumbbells is governed by
\begin{eqnarray}
\boldsymbol{u}_t + \boldsymbol{u}\cdot\boldsymbol{\nabla}\boldsymbol{u} = -\boldsymbol{\nabla} p + {\Rey}^{-1}\left[{\beta}\nabla^2\boldsymbol{u} + {(1-\beta)}\boldsymbol{\nabla}\cdot\boldsymbol{\tau}\right], \label{gov_eqn_u} \\
\textbf{\textit{\textsf{c}}}_t + \boldsymbol{u}\cdot\boldsymbol{\nabla}\textbf{\textit{\textsf{c}}} - \textbf{\textit{\textsf{c}}}\cdot\boldsymbol{\nabla}\boldsymbol{u} -(\boldsymbol{\nabla}\boldsymbol{u})^{\rm T}\cdot\textbf{\textit{\textsf{c}}} = -\boldsymbol{\tau}, \label{gov_eqn_c} 
\end{eqnarray}
where the subscript ``$t$" refers to the time derivative, $\boldsymbol{\tau} = (f\textbf{\textit{\textsf{c}}}-\textbf{\textit{\textsf{I}}})/W$ is the elastic stress of polymer, $\beta = \mu_s/\mu$, $W = \lambda U_c/R$, and $Re = \rho U_c R/\mu$ is the Reynolds number. 
The quantities $\mu_s$ and $\mu$ refer to the solvent viscosity and total viscosity, respectively, while $U_c$ is the centreline velocity when no polymer is added. In addition, $R$ and $\lambda$ refer to the pipe radius and relaxation time, respectively. In Eq.~\eqref{gov_eqn_c}, the conformation tensor is $c_{ij}=\langle \tilde{R}_i \tilde{R}_j \rangle$, where $\tilde{R}_i$ classically represents the end-to-end vector of the polymer molecule, and $\textbf{\textit{\textsf{I}}}$ is the identity matrix. The function $f$ in the definition of $\boldsymbol{\tau}$ is given by
\begin{equation}
f = \frac{L^2-3}{L^2-\text{tr}(\textbf{\textit{\textsf{c}}})}, \label{eq:pet_func}
\end{equation}
and is the Peterlin function under the FENE-P model, 
{
and serves as a restoring tendency like the spring constant in Hooke's law, except that $f$ is a functions of $\textbf{\textit{\textsf{c}}}$ implying the nonlinear dependence of the polymer stress on the strain. The parameter, $L$($>\sqrt{3}$) is the mean length of extensibility of polymer molecules. The $f$ defined here has a correction of `$-3$' in the numerator following Refs.~\cite{beris1994thermodynamics,dimitropoulos1998direct}.
}

\begin{figure}\centering
	\begin{overpic}[width=0.40\textwidth]{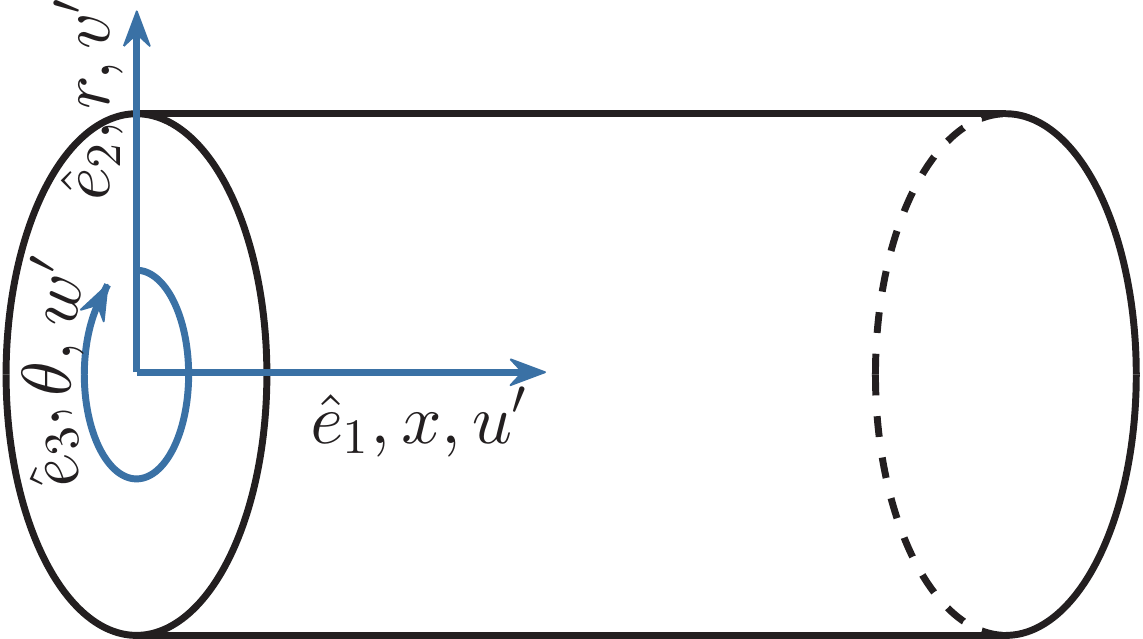}
	\end{overpic}
	\caption{Schematic diagram of the flow set-up with associated notations in cylindrical coordinates\label{fig:sketch}}
\end{figure}

Lastly, the velocity field $\boldsymbol{u} = (u, v, w)^{\rm T}$ is considered in cylindrical coordinates $(x,r,\theta)$ with axial ($\boldsymbol{\hat{e}_1}$), radial ($\boldsymbol{\hat{e}_2}$) and 
azimuthal ($\boldsymbol{\hat{e}_3}$) directions, respectively (see Fig~\ref{fig:sketch}).
With these notations, the divergence of the elastic stress in Eq.~\eqref{gov_eqn_u} is given by $\boldsymbol{\nabla} \cdot \boldsymbol{\tau} = [\tau_{11x} + \tau_{21r} + (\tau_{21}+\tau_{31\theta})/r]\boldsymbol{\hat{e}_1} + [\tau_{12x} + \tau_{22r} + (\tau_{22} - \tau_{33}+\tau_{32\theta})/r]\boldsymbol{\hat{e}_2} + [\tau_{13x} + \tau_{23r} + (\tau_{23} + \tau_{32}+\tau_{33\theta})/r]\boldsymbol{\hat{e}_3}$. The suffixes $x$, $r$ and $\theta$ imply differentiation with respect to these variables.

First, it is worth highlighting that in the steady laminar case, with $\textbf{\textit{\textsf{c}}} = \textbf{\textit{\textsf{C}}}(r)$, $\boldsymbol{u}= \boldsymbol{U}(r) = [U(r), 0, 0]$, $p = P(x) = -4x/\Rey$ and  $F(r)=(L^2-3)/(L^2-\text{tr}(\textbf{\textit{\textsf{C}}}))$.  The solution of Eqs.~\eqref{gov_eqn_u}--\eqref{gov_eqn_c} is given by:
\begin{align}
F(r) &= 1+(\zeta_1^{1/3} + \zeta_2^{1/3} -2)/(3\beta), \label{eq_F} \\
U(r) &= 2\int_0^1 \frac{r'F(r')[1-\mathcal{U}(r-r')]}{\beta [F(r')-1]+1}\diff r', \label{U_eqn} \\
C_{11}(r) &=(2W^2U_r^2+F^2)/F^3, \label{eq_C11}  \\
C_{22}(r) &= C_{33}(r) = 1/F, \label{eq_C22C33}  \\
C_{12}(r) &= W \,U_r/F^2, \label{eq_C12} 
\end{align}
where $\zeta_1 = a + \sqrt{a^2-1}$, $\zeta_2 = a - \sqrt{a^2 -1}$ and $a = 1 + 108\beta W^2r^2/L^2$ with positive square-root and real cubic-root implied~\cite{malik2020viscoelastic,cruz2005analytical}. Note that $\mathcal{U}(r)$ denotes the Heaviside step function, and that all functions in Eqs.~\eqref{eq_F}--\eqref{eq_C12} hold even or odd parities with respect to $r$.  Specifically, $C_{12}$ has odd parity while $F, U, C_{11}, C_{22}$ and $C_{33}$ have even parities as shown by Malik {\it et al.}~\cite{malik2020viscoelastic}. 

However, these parity properties can also be obtained from the consistency arguments under the transformation, $(r,\theta)\rightarrow (-r, \theta \pm \pi)$, where $\theta$ is the azimuthal coordinate. Such arguments have been used by Lewis and Bellan~\cite{lewis1990physical} for scalars and vectors (see also Refs.~\cite{kerswell1996linear,mohseni2000numerical} for applications). In the case of a dyadic such as $\textbf{\textit{\textsf{C}}}(r)$ as noted by R. R. Kerswell (personal communication, August 14, 2020), similar analysis after noting the two unit vectors in $\textbf{\textit{\textsf{C}}}(r)$ deduce that $C_{11}, C_{22}$ and $C_{33}$ are even and $C_{12}$ is odd (See also the appendix of~\cite{kerswell1996linear}).

These properties of the mean flow variables will be used to obtain the symmetries and leading-order behaviours of the perturbations in what follows. For completeness, it is worth mentioning that an alternate formulation of the FENE-P model, its laminar mean profiles, and its relation to the Phan-Thien and Tanner model can be found in Cruz {\it et al.}~\cite{cruz2005analytical} and Poole~\cite{poole2019similarities}. The pioneering solution for the mean-flow for an inviscid solvent was obtained by Oliveira~\cite{oliveira2002exact}.

\section{Structure of small perturbations}\label{sec:linsys}
Let the mean-state, $(\boldsymbol{U}, P, \textbf{\textit{\textsf{C}}})^{\rm T}$ be perturbed with $(\boldsymbol{u'}, p', \textbf{\textit{\textsf{c'}}})^{\rm T} \exp[\mi(\alpha x + n\theta - \omega t)]$ where $n\in\mathbb{Z}$, $\alpha\in\mathbb{R}$, and $\omega\in\mathbb{C}$. Let $G \equiv \partial F/\partial[\text{tr}(\textbf{\textit{\textsf{C}}})]$, i.e., $G = 
(L^2-3)/(L^2-\text{tr}(\textbf{\textit{\textsf{C}}}))^2$ and $\Gamma = (1-\beta)/(ReW)$. The amplitude of the fluctuating part of $f$ is given by $f' = \text{tr}(\textbf{\textit{\textsf{c'}}})G$. The linearised equations for small perturbations read
\begin{align}
\mathcal{L}u'=&-U_rv'-\mi\alpha p' +\Gamma\left [ (G_1 +\mi\alpha F)c_{11}' + G_1(c_{22}' + c_{33}') + C_{12}GD(c_{11}'+c_{22}'+c_{33}') \right . \nonumber \\
&\left .  + (F_r + Fr^{-1} + FD)c_{12}' + \mi n Fr^{-1}c_{13}'\right ], \label{eq:umom}\\
\mathcal{L}v'=&-p'_r - \beta(r^2\Rey)^{-1}\left(v'+\mi 2 nw'\right) +\Gamma\left [ G_2c_{11}' + (G_2+F_r+Fr^{-1})c_{22}' + \mi\alpha Fc_{12}'  \right . \nonumber \\
&\left .  + (G_2-Fr^{-1})c_{33}' + C_{22}GD(c_{11}'+c_{33}') +(F+C_{22}G)Dc_{22}' + \mi n Fr^{-1}c_{23}'  \right ],
\label{eq:vmom}\\
\mathcal{L}w'=&-\mi n r^{-1}p' - \beta(r^2\Rey)^{-1}\left(w'-\mi 2 nv'\right) +\Gamma\left [\mi \alpha F c_{13}' +(F_r + 2Fr^{-1}+F D)c_{23}' \right . \nonumber \\
&\left . + \mi n C_{33} G r^{-1}(c_{11}'+ c_{22}')  + \mi n r^{-1}(F+C_{33}G)c_{33}'\right ],\label{eq:wmom}\\
G_3c_{11}' =& -C_{11r}v' + 2(\mi \alpha C_{11}u' + C_{12}Du' + U_r c_{12}') -C_{11}GW^{-1}(c_{11}'+ c_{22}'+ c_{33}'),\label{eq:c11t}\\
G_3c_{22}' =& ( 2\mi \alpha C_{12} -C_{22r} + 2C_{22}D )v' -C_{22}GW^{-1}(c_{11}'+ c_{22}'+ c_{33}'),\label{eq:c22t}\\
G_3c_{33}' =& (2C_{33}r^{-1}-C_{33r})v' + 2\mi n C_{33}r^{-1}w' -C_{33}GW^{-1}(c_{11}'+ c_{22}'+ c_{33}'),\label{eq:c33t}\\
G_3c_{12}' =& (\mi \alpha C_{11}-C_{12r}+ C_{12}D)v' + (\mi\alpha C_{12}+ C_{22}D)u' + (U_r - C_{12}GW^{-1})c_{22}'  \nonumber \\
&   -C_{12}GW^{-1}(c_{11}'+ c_{33}'),\label{eq:c12t}\\
G_3c_{13}' =& (\mi \alpha C_{11}-C_{12}r^{-1}+ C_{12}D)w' + U_rc_{23}' + \mi n C_{33}r^{-1}u' ,\label{eq:c13t}\\
G_3c_{23}' =& (\mi \alpha C_{12}-C_{33}r^{-1}+ C_{22}D)w' + \mi n C_{33}r^{-1}v', \label{eq:c23t}
\end{align}
where $\mathcal{L} = \mi(\alpha U -\omega ) - \beta\Rey^{-1}\left [ D^2 + 
r^{-1}D -\left  (\alpha^2+n^2r^{-2}\right ) \right ]$, $D = \frac{\mathrm{d}}{\mathrm{d}r}$, $G_1(r) = 
(\mi \alpha C_{11}+ C_{12r} + C_{12}r^{-1})G + C_{12}G_r$, $G_2(r) = (\mi \alpha C_{12}+ C_{22r})G + 
C_{22}G_r$ and $G_3(r) = \mi(\alpha U -\omega )+FW^{-1}$. As a matter of convention, we use the suffix $r$ to represent the derivative with respect to $r$ of mean flow variables, and operator $D$ to imply the same for the fluctuations.

\subsection{Analytic behaviour}
\label{sec:analyticbehav}
As can be noted from the operator $\mathcal{L}$, Eqs.~\eqref{eq:umom}--\eqref{eq:wmom} exhibit a singularity at $r=0$. Hence, regularity conditions are required to preclude non-analytic solutions. If each term of Eqs.~\eqref{eq:umom}--\eqref{eq:c23t} goes like $\sim r^j$, where $j\in\mathbb{Z}^+$ can be different for each term, then in the limit of $r \rightarrow 0$, these equations would become redundant as each of them would have factors of zeros. However, Eqs. \eqref{eq:umom}--\eqref{eq:c23t} can convey useful information in the limit of $r \rightarrow 0$ if the greatest common factor, say $r^i$, of all the terms is factored out, thus serving to enforce the boundary conditions at the pipe centre~\cite{malik2019linear}. To get the actual value of this integer $i$, the behaviour of the unknowns in this limit is indispensable and can be derived by means of a Taylor analysis around $r=0$.

Let $\rchi_j(r)$ with $j = 0,\cdots,9$ be analytic with Taylor expansions, $\sum_{k=0}^\infty r^k\rchi_{j,k}$ for all $j$. We use these family of functions to represent parts of velocities, pressure and conformation tensors as shown below. A Taylor analysis of Eqs.~\eqref{eq:umom}--\eqref{eq:c23t} shows that the solutions take the form
\begin{align}
(p', u', v', w') =& \left \{
\begin{array}{ll}
(\rchi_0, \rchi_1, r\rchi_2, r\rchi_3) &\hspace{1.7cm}\mbox{for}  \ n = 0,\\
(r^{|n|}\rchi_0, r^{|n|}\rchi_1,r^{|n|-1}\rchi_2,r^{|n|-1}\rchi_3) &\hspace{1.7cm}\mbox{for}   \ n \neq 0,\\
\end{array}
\right . \label{eq:vel_behaviour}
\\
(c_{11}',c_{22}',c_{33}') =& \left \{ 
\begin{array}{ll}
(\rchi_4, \rchi_5, \rchi_6)  \hspace{3.6cm} \text{for} \ \alpha \neq 0 & \text{and} \ n = 0, \\
(r^2\rchi_4, r^2\rchi_5, r^2\rchi_6)  \hspace{2.6cm}\text{for} \ \alpha = 0 & \text{and} \ n = 0, \\
(r\rchi_4, r\rchi_5, r\rchi_6) & \text{for} \  |n| = 1, \\
(r^{|n|}\rchi_4, r^{|n|-2}\rchi_5, r^{|n|-2}\rchi_6) & \text{for} \  |n| \geq 2,\\
\end{array}
\right .\label{eq:cii_behaviour}
\\
(c_{12}',c_{13}',c_{23}') =& \left \{ 
\begin{array}{ll}
(r\rchi_7, r\rchi_8, r^2\rchi_9) \hspace{2.9cm} \text{for} \ \alpha \neq 0 & \text{and} \  n = 0, \\
(r\rchi_7, r^3\rchi_8, r^2\rchi_9) \hspace{2.7cm}  \text{for} \ \alpha = 0 & \text{and} \  n = 0, \\
(\rchi_7, \rchi_8, r\rchi_9) & \text{for} \  |n| = 1, \\
(r^{|n|-1}\rchi_7,r^{|n|-1}\rchi_8, r^{|n|-2}\rchi_9) & \text{for} \  |n| \geq 2, \\
\end{array}
\right .\label{eq:cij_behaviour}
\end{align}
and that the functions $\rchi_j(r)$ are even functions of $r$, and $\rchi_2(r) = 0$ for $\alpha =0$. The detailed derivation of Eqs.~\eqref{eq:vel_behaviour}--\eqref{eq:cij_behaviour} is given in the appendices~\ref{appA}-\ref{appD} by the method of Frobenius. However, these solution forms can also be arrived at via couple of other methods. A heuristic method is given in the next subsection. As an observation, it should be noted that the behaviours of $u', v',\ \text{and}\ w'$ shown in Eq.~\eqref{eq:vel_behaviour} are same as those for the Newtonian case~\cite{priymak1998accurate,meseguer2003linearized,malik2019linear}. 

As a third way for arriving at the ansatz in Eqs.~(\ref{eq:vel_behaviour})-(\ref{eq:cij_behaviour}), one can follow the method of Fourier analysis~\cite{lewis1990physical}. This analysis predicts the leading order power law behaviour and the parities of the scalars such as $p'$ and the vectors in the plane perpendicular to polar axis such as $(v', w')$ irrespective of the physical problems at hand. Thereafter, the behaviours of $u'$ and the components of $\textbf{\textit{\textsf{c'}}}$ can be obtained from the continuity equation, and the governing equations for \textbf{\textit{\textsf{c'}}}, i.e., Eqs~(\ref{eq:c11t})-(\ref{eq:c23t}). In Appendix~\ref{appE}, we show this for the case of $(n=0; \alpha\neq 0)$ and $(n=0; \alpha= 0)$. The other cases can be handled in the same way to arrive at the ansatz in Eqs.~(\ref{eq:vel_behaviour})-(\ref{eq:cij_behaviour}).

\subsection{Heuristic derivation \label{sec:heur}} 
Here we present a heuristic approach to derive the ansatzes in Eqs.~\eqref{eq:vel_behaviour}--\eqref{eq:cij_behaviour}. Let us first consider the limit of $\beta \rightarrow 1$ where the contributions due to polymer stress in Eqs.~\eqref{eq:umom}--\eqref{eq:wmom} are identically null. The case of $\beta \neq 1$ is addressed at the end of this subsection. 

In this limit of $\beta \rightarrow 1$, the velocities in Eq.~\eqref{eq:vel_behaviour} acquire the behaviour of the Newtonian flow. The behaviours of $c_{ij}'$ are determined by $(u', v', w')$ according to Eqs.~\eqref{eq:c11t}--\eqref{eq:c23t}. First, let us consider the case $n = 0$. We take into consideration the parities of the mean flow variables shown at the end of Sec.~\ref{sec:model}. These imply that $C_{11}$, $C_{22}$, $C_{33}$, $G$, $G_1$ and $G_3$ are even functions with constant-like behaviours as $r \rightarrow 0$, and that $U_r$, $C_{12}$ and $G_2$ are odd and linear to the leading order. From Eq.~\eqref{eq:vel_behaviour}, we have $w' = r\rchi_3$. Therefore, the terms $C_{22}Dw' - r^{-1}C_{33}w'$ and $\mi \alpha C_{12}w'$ in the RHS of Eq.~\eqref{eq:c23t} goes like $r^2$ times an even function in the limit $r\rightarrow 0$. Since $G_3$ is even, Eq.~\eqref{eq:c23t} implies $c_{23}'= r^2\rchi_9$ as given by Eq.~\eqref{eq:cij_behaviour}. Similarly, Eq.~\eqref{eq:c23t} suggests that $c_{13}'\sim r$ multiplied by an even function of $r$ to the leading order, hence the definition $c_{13}'= r\rchi_8$ in Eq.~\eqref{eq:cij_behaviour}. Similar analyses establish the behaviours of $c_{11}'$, $c_{22}'$, $c_{33}'$ and $c_{12}'$ as in Eqs.~\eqref{eq:cii_behaviour}--\eqref{eq:cij_behaviour} although their equations are coupled. It is worth adding that these coupled equations can be solved algebraically for each of the variables and analysed individually. Here, only three variables are independent, namely $c_{11}'+c_{33}'$, $c_{22}'$ and $c_{12}'$. In addition, the equation for $c_{11}'+c_{33}'$ can be formed by adding Eq.~\eqref{eq:c11t} and Eq.~\eqref{eq:c33t}. 

Now, let us consider the case $|n| = 1$ for $\beta \rightarrow 1$. A similar analysis can be carried out after noting the following element. The terms $2C_{33}(v'+n \mi w')/r$ and $2C_{33}(-w' +n \mi v')/r$ on the RHS of Eq.~\eqref{eq:c33t} and Eq.~\eqref{eq:c23t}, respectively, are of $O(r)$, since $v'+n\mi w' \sim r^2$. This can be derived using the fact that Khorrami {\it et al.}~\cite{khorrami1989application} have shown that $v'+ n\mi w' =0$ at the centreline for $|n|=1$. Since $v'$ and $w'$ are even functions from Eq.~\eqref{eq:vel_behaviour} for $|n|=1$, $(v'+ n \mi w')_{r=0}=0$ implies $v'+ n\mi w' \sim r^2$. While analysing the terms for $n=\pm 1$, one should note that $(c_{12}'+n\mi c_{13}')_{r=0} = 0$ from Eq.~\eqref{eq:c12t} and Eq.~\eqref{eq:c13t}.

Further for $|n|=1$, the term $F(c_{12}'+n\mi c_{13}')/r$ in Eq.~(\ref{eq:umom}) deserves attention. One should note that $(c_{12}'+n\mi c_{13}')_{r=0} = 0$ from Eq.~\eqref{eq:c12t} and Eq.~\eqref{eq:c13t}. Therefore the arguments in the above paragraph for $(v'+n \mi w')/r$ applies, hence $(c_{12}'+n\mi c_{13}')/r = 0$ as $r\rightarrow 0$.

For $n\geq 2$, all terms of Eqs.~\eqref{eq:c11t}--\eqref{eq:c23t} remain, and the analysis is essentially the same as for the case $n=0$. Finally, let us consider the case $\beta \neq 1$, whereby there is a non-trivial contribution from the polymer stress terms in Eqs.~\eqref{eq:umom}--\eqref{eq:wmom}. It should be noted that the functional forms of the solution set given by Eqs.~\eqref{eq:vel_behaviour}--\eqref{eq:cij_behaviour} will still remain valid, as long as the parity of each of these polymer terms in Eqs.~\eqref{eq:umom}--\eqref{eq:wmom} conforms with the parities of other terms of these equations for each $n$. If the parities are indeed the same, the polymer terms only amount to a modification in the multiplicative constants in the Taylor expansion of the other terms in Eqs.~\eqref{eq:umom}--\eqref{eq:wmom} allowing the velocity field to preserve the form as in Eq.~\eqref{eq:vel_behaviour}. This, in turn, implies the preservation of Eqs.~\eqref{eq:cii_behaviour}--\eqref{eq:cij_behaviour} by Eqs.~\eqref{eq:c11t}--\eqref{eq:c23t}. We find that that turns out to be the case. With the parities of the mean flow variables taken into consideration, every term of the polymer stress is in perfect `harmony' with other terms. 

\section{Numerical validation \label{sec:valid}}
Now, we verify these ansatzes in Eq.~(\ref{eq:cii_behaviour}) and Eq.~(\ref{eq:cij_behaviour}) numerically. For a general value of $\beta$, such ansatzes are necessary for accurate computation of the solutions with proper specification of regularity. On the other hand, numerical results from any method that utilizes these ansatzes would not serve as a verification for the very same. 

\begin{figure}\centering
	\includegraphics[width=0.8\textwidth]{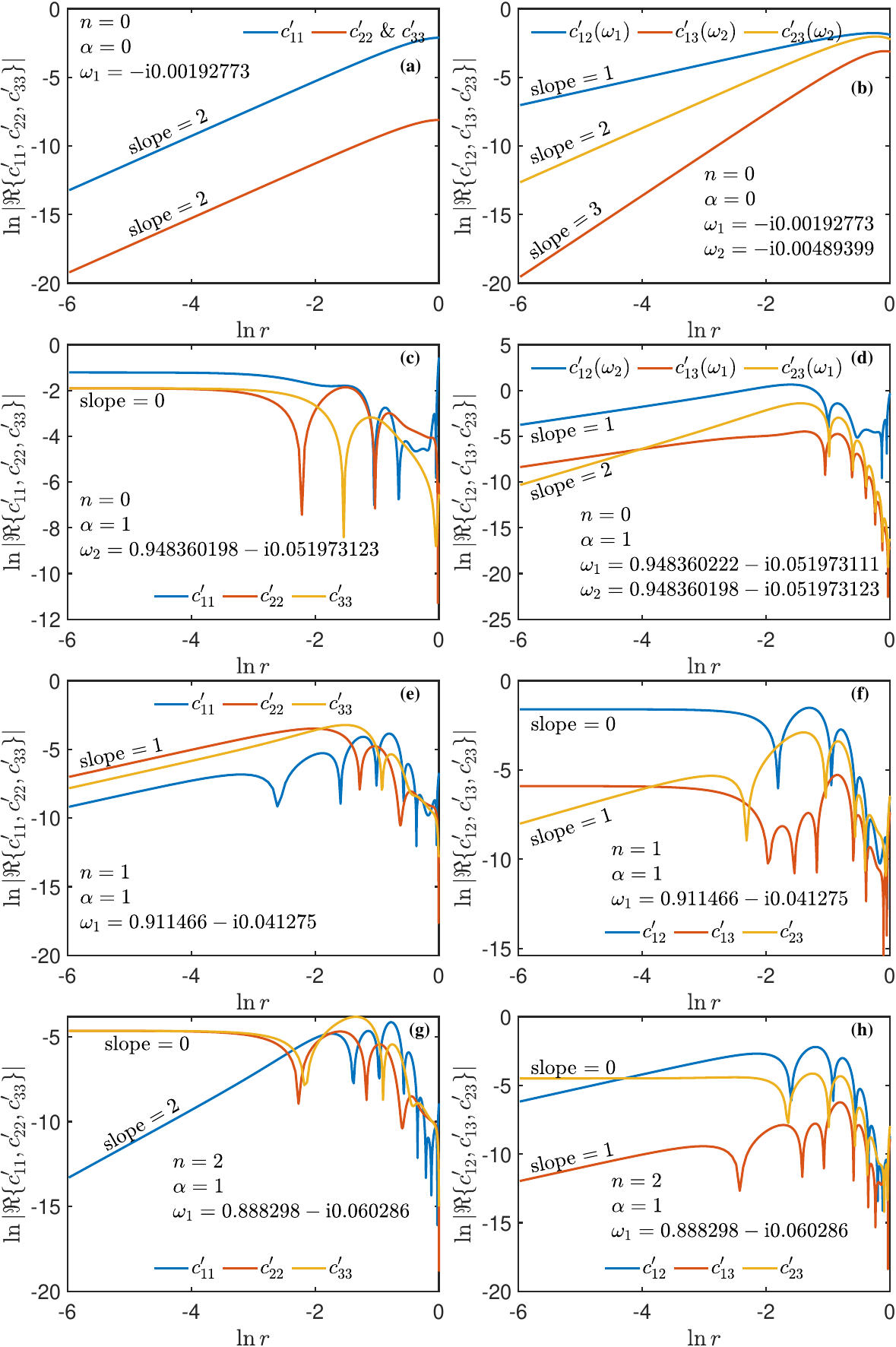}
	\caption{The components of $\textbf{\textit{\textsf{c'}}}$ of the least decaying modes arising from solvent for $Re = 3000$, $W = 0.1$, $L = 20$ and $\beta = 1$: (left) diagonal elements of $\textbf{\textit{\textsf{c'}}}$; (right) off-diagonal elements of $\textbf{\textit{\textsf{c'}}}$. In panels (a)-(d), $\omega_1$ and $\omega_2$ refer to the first and second least-decaying mode, respectively. \label{fig:loglog}}
\end{figure}

However, for $\beta = 1$, the eigenvalue problem of the momentum Eqs.~(\ref{eq:umom})-(\ref{eq:wmom}) decouples from that of the components, $c_{ij}'$\textsc{\char13}s of the conformation tensor formed by Eqs.~(\ref{eq:c11t})--(\ref{eq:c23t}), and the velocity perturbations are completely determined by the flow of the pure solvent which is Newtonian. Therefore, this gives a window of opportunity to verify these ansatzes for $c_{ij}'$ with velocity perturbations, $u'$, $v'$ and $w'$, and the corresponding eigenvalues specified from a Newtonian stability code. For this purpose, we use the code given in the appendix of Malik {\it et al.}~\cite{malik2019linear}. The corresponding $c_{ij}'$ are given by
\begin{equation}
\{c_{11}',c_{22}',c_{33}',c_{12}',c_{13}',c_{23}'\}^{\rm T} = (A+G_3I)^{-1}b, \label{eq:c_beta_1}
\end{equation}
which is obtained from Eqs.~(\ref{eq:c11t})--(\ref{eq:c23t}). The non-zero elements of the $6\times 6$ matrix $A$ are: $A_{11}\! = \!A_{12} \!=\! A_{13} \!=\! C_{11}G/W$, $A_{14} \!=\! -2U_r$, $A_{21} \!=\! A_{22} \!=\! A_{23} \!=\! C_{22}G/W$, $A_{31} \!=\! A_{32} \!=\! A_{33} \!=\! C_{33}G/W$, $A_{41} \!=\! A_{43} \!=\! C_{12}G/W$, $A_{42} \!=\! (C_{12}G/W)-U_r$, $A_{56} \!=\! -U_r$. The elements of the column vector $b$ are given by
\begin{align}
b_1 =& -C_{11r}v' + 2\mi \alpha C_{11}u' + 2C_{12}Du' ,\\
b_2 =& ( 2\mi \alpha C_{12} -C_{22r} + 2C_{22}D )v',\\
b_3 =& (2C_{33}r^{-1}-C_{33r})v' + 2\mi n C_{33}r^{-1}w',\\
b_4 =& (\mi \alpha C_{11}-C_{12r}+ C_{12}D)v' + (\mi\alpha C_{12}+ C_{22}D)u',\\
b_5 =& (\mi \alpha C_{11}-C_{12}r^{-1}+ C_{12}D)w' + \mi n C_{33}r^{-1}u' ,\\
b_6 =& (\mi \alpha C_{12}-C_{33}r^{-1}+ C_{22}D)w' + \mi n C_{33}r^{-1}v'.
\end{align}
The $c_{ij}'$s obtained with parameters, $(n,\alpha) \in \{(0,0), (0,1), (1,1), (2,1)\}$, $\beta = 1$, $Re = 3000$, $W = 0.1$ and $L=20$ for the least-decaying modes of momentum equations are shown in Fig.~\ref{fig:loglog}. 

For the cases of $n = 0$, shown in Fig.~\ref{fig:loglog}(a)--(d), we consider the first two least-decaying modes (denoted as $\omega_1$ and $\omega_2$ in Fig.~\ref{fig:loglog}), since some of the $c_{ij}'$ components are zeros if we consider only the least-decaying mode. For example, at $(n,\alpha) = (0,0)$, the components $c_{13}' = c_{23}'=0$ for the first least-decaying mode with $\omega = -{\rm i}0.00192733$. Therefore, we also consider the second least-decaying mode with $\omega = -{\rm i}0.00489399$ for the components $c_{13}'$ and $c_{23}'$.

For $n = 1$ and $2$, we consider only one mode to present the components of $c_{ij}'$. 
In Fig.~\ref{fig:loglog},  $\ln |\Re (c_{ij}')|$'s have been plotted against $\ln r$ in order to bring the exponents as the slope of the curve in the limit $r\rightarrow 0$ (i.e., $\ln r \rightarrow $ large negative values). The symbol $\Re$ stands for the {\rm real part} of its argument.

As can be noted from each panel in Fig.~\ref{fig:loglog}, the the power-law behaviours of $c_{ij}'$\textsc{\char13}s shown of Eq.~(\ref{eq:cii_behaviour}) and Eq.~(\ref{eq:cij_behaviour}) are verified numerically, since their leading-order exponents have translated into their respective slopes in these log-log plots. 

We have also verified the parities of $\rchi_j(r)$'s by performing successive differentiations of these $\rchi_j(r)$'s---obtained from dividing $c_{ij}'(r)$'s by the corresponding leading order power-law behaviour in Eq.~(\ref{eq:cii_behaviour}) and Eq.~(\ref{eq:cij_behaviour})---using Chebyshev differentiation matrices, and by noticing that $D^k\rchi_j(r) \rightarrow 0$ as $r\rightarrow 0$ for every odd $k^{\rm th}$ order of differentiations (not shown in Figs.). 

\section{Physical significance and further discussions \label{sec:signific}} 
The physical significance of the result, $(c_{12}'+n\mi c_{13}')_{r=0} = 0$ for $|n| = 1$ is that $\lim_{r \rightarrow 0}[(c_{1j}'\boldsymbol{\hat{e}_1\hat{e}_j})_\theta]=0$. This is similar to the condition $\lim_{r \rightarrow 0}[(\boldsymbol{u'})_\theta]=0$ that results in $(v' + n\mi w')_{r=0} =0$ as found by Khorrami {\it et al.}~\cite{khorrami1989application} for the Newtonian case and applicable to the present non-Newtonian flow. The condition $\lim_{r \rightarrow 0}[(c_{1j}'\boldsymbol{\hat{e}_1\hat{e}_j})_\theta]=0$ implies that the traction vector along $\boldsymbol{\hat{e}_1}$ is uniform along $\boldsymbol{\hat{e}_3}$ as we approach the centreline. As evident from Eq.~\eqref{eq:c12t} and Eq.~\eqref{eq:c13t}, $c_{12}'$ and $c_{13}'$ at the centreline are due to $v'$ and $w'$, respectively, to the leading order. This implies a shearing of polymers against the restoring tendency by $v'$ and $w'$, but without stretching, given that the stretching terms are of higher orders. Since $(v' + n\mi w')_{r=0} =0$, this results in $(c_{12}'+n\mi c_{13}')_{r=0} = 0$ as a consequence. In fact, even the mean $C_{12}$ arises due to the restoring tendency and mean fluid shear, $U_r$, and results in shear thinning (see Ref.~\cite{malik2020viscoelastic} for more details).

In another way of looking at this result for $|n|=1$, one should note that the axial traction vector  $c_{1j}'\boldsymbol{\hat{e}_1\hat{e}_j}$ (apart from the factor of $F$ that is neglected here), is predominantly in the plane perpendicular to the axis, since $c_{11}'$ is of higher order than $c_{12}'$ and $c_{13}'$. Therefore, the the general result for such vectors from Fourier analysis by Lewis and Bellan~\cite{lewis1990physical} apply, which would predict the current observation, i.e., $(c_{12}'+n\mi c_{13}')_{r=0} = 0$. 

Further, the components of the axial traction, $c_{1j}'\boldsymbol{\hat{e}_1\hat{e}_j}$ have the same behaviour as the velocity vector $(u', v', w')$ for all values of $n$ as long as $\alpha \neq 0$. This indicates the close relation between the axial traction and the velocities, where the components of the former inherits these properties from the latter via Eqs.~\ref{eq:c11t}-\ref{eq:c13t}.

Eq.~\ref{eq:cii_behaviour} suggests that for $|n|\geq 2$, the stretching of the polymer molecules in the plane perpendicular to the axial direction is dominant over the same along the axial direction close to the centreline. The particular value of $|n|=2$ deserves attention, since the components $c_{22}'$ and $c_{33}'$ do not vanish at the centre. However, the sum, $c_{22}'+ c_{33}'$ indeed vanishes, which implies that, under these linear level dynamics, the molecules are in random rotational motion in the plane perpendicular to the polar axis around $r=0$. Further, we note that the trace vanishes near centreline for all values of $n$ which implies that the perturbation do not cause any fluctuations in the polymer length at the centreline.

The fact that the sum $c_{22}'+ c_{33}'\rightarrow 0$ while $c_{22}$ and $c_{33}$ remain finite for $n = 2$ demonstrates that the fluctuation $\textbf{\textit{\textsf{c'}}}$ may not be positive definite away from the centreline. The Refs.~\cite{hameduddin2018geometric,hameduddin2019perturbative} focuses on the issue of ensuring that the total conformation tensor, $\textbf{\textit{\textsf{c}}} = \textbf{\textit{\textsf{C}}} + \textbf{\textit{\textsf{c'}}}$ remains positive definite, a requirement for avoiding unphysical negative value for polymer length.

\section{Polymers' natural, continuous and stratified modes \label{sec:natural}}
At $\beta=1$, there are other modes arising from Eq.~(\ref{eq:c_beta_1}). These are the {\it natural} modes of the system that exist for the trivial solutions for velocity perturbations, i.e., $u'(r) = v'(r) = w'(r) = 0$, and they satisfy the condition
\begin{equation}
\det(A + G_3I) = 0. \label{eq:det}
\end{equation}
These modes are independent of $n$, so the power-law behaviour predicted in Eq.~(\ref{eq:cii_behaviour}) and Eq.~(\ref{eq:cij_behaviour}) do not apply. The natural modes satisfy the system,
\begin{equation}
(A + G_3I)\{c_{11}',c_{22}',c_{33}',c_{12}',c_{13}',c_{23}'\}^{\rm T} = 0,
\end{equation}
which shows that $c_{13}' = c_{23}' = 0$ and $c_{22}' = c_{33}'$, hence reducing the system to two dimensions. Therefore, the reduced system can be written in terms of three unknowns as shown below:
\begin{equation}
\begin{pmatrix}
\alpha U\!-{\rm i}(F\!+\!C_{11}G)/W & -2{\rm i}C_{11}G/W & 2{\rm i}U_r\\
-{\rm i}C_{22}G/W & \alpha U\!-{\rm i}(F\!+\!2C_{22}G)/W & 0\\
-{\rm i}C_{12}G/W & -2{\rm i}C_{12}GW^{-1}+{\rm i}U_r  & \alpha U\!-{\rm i}\!F/W
\end{pmatrix}
\begin{pmatrix}
c_{11}'\\
c_{22}'\\
c_{12}'
\end{pmatrix}
= \omega
\begin{pmatrix}
c_{11}'\\
c_{22}'\\
c_{12}'
\end{pmatrix}. \label{eq:natural}
\end{equation}
\begin{figure}\centering
	\includegraphics[width=0.6\textwidth]{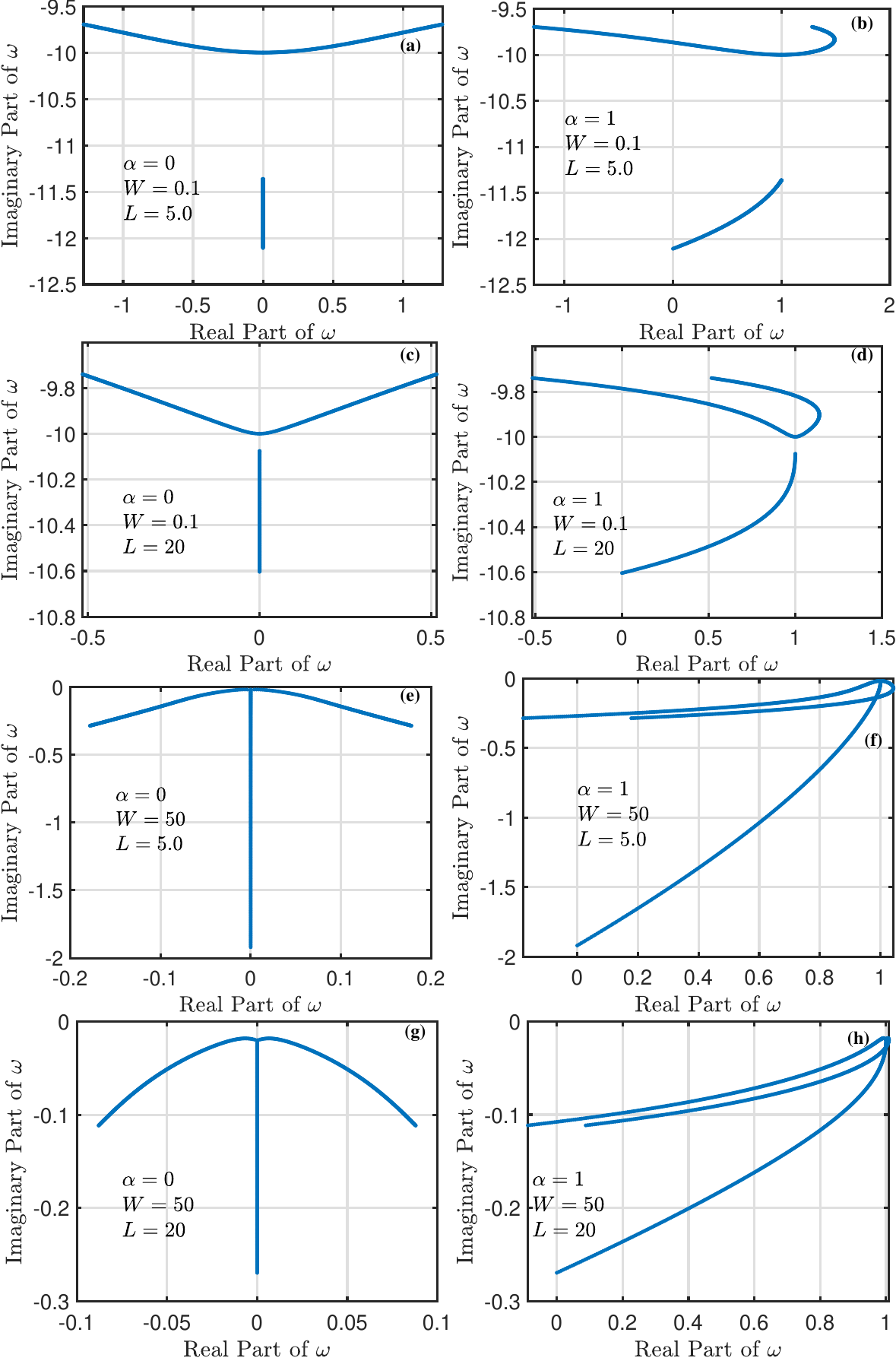}
	\caption{The spectrum of the natural modes of polymer conformation tensor from Eq.~(\ref{eq:natural}): (left) $\alpha = 0$; (right) $\alpha = 1$ \label{fig:naturalmodes}}
\end{figure}
\begin{figure}\centering
	\includegraphics[width=0.6\textwidth]{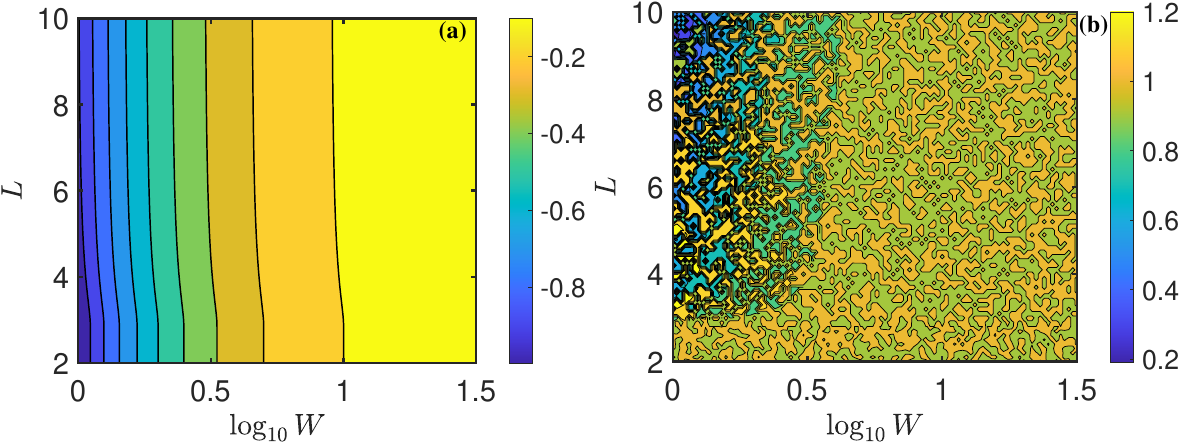}
	\caption{The contours of growth/decay rates and phase speeds of the least-decaying mode from Eq.~(\ref{eq:natural}) in the $(W,L)$ plane for $\alpha = 1$: (a) the growth/decay rates; (b) the phase speed \label{fig:natmode_lstdcy_omgr_omgi}}
\end{figure}
The spectrum obtained from Eq.~(\ref{eq:natural}) using QZ algorithm is shown in Fig.~\ref{fig:naturalmodes}, which reveals a continuous behaviour. The eigenfunctions do not have any other imposed conditions to be satisfied at the wall, a fact that leads to a continuous spectrum from the spectral theory. 

Figures~\ref{fig:naturalmodes}(e)-(h) present the spectrum for $W = 0.1$ for two values of $\alpha$ and $L$. Upon comparing the panels Fig.~\ref{fig:naturalmodes}(a) against Fig.~\ref{fig:naturalmodes}(c), or Fig.~\ref{fig:naturalmodes}(b) against Fig.~\ref{fig:naturalmodes}(d), an increase in $L$ stabilizes the modes further by enhancing the decay rates. This phenomenon is observed irrespective of the value of $W$. Figures~\ref{fig:naturalmodes}(e)-(h) present the spectrum for parameters same as Figs.~\ref{fig:naturalmodes}(a)-(d) but for an increased $W$($=50$), where the stabilizing role of an increasing $L$ is shown. Since $L$ does not appear in the system~(\ref{eq:natural}) explicitly, it affects the decay rates only due to change in the mean-flow variables. 

Similarly, upon comparing, for example, Fig.~\ref{fig:naturalmodes}(a) against Fig.~\ref{fig:naturalmodes}(e), we find that an increase in $W$ destabilizes the system by decreasing the decay rate. However, we observed that a further increase in $W$ only takes the least-decaying mode towards marginal stability (i.e., $\lim_{W\rightarrow \infty} \Im(\omega)\rightarrow 0^-$).

Unlike the cases of viscous flow, a change in $\alpha$ does not alter the growth rate of the least-decaying modes significantly (compare the panels on the right column of Fig.~\ref{fig:naturalmodes} to that on the left). A non-zero value of $\alpha$ causes the majority of modes to travel in the direction of the mean flow. This is tied to the fact that the $\alpha$ appears only as a scaling to the mean-flow velocity in the system~(\ref{eq:natural}). It is well known from modal energy analyses that a convective term cannot contribute to the energy growth since they will only redistribute energy with in the domain (see for example {Malik et al.}~\cite{malik2018growth,malik2006nonmodal}).

Since $\alpha$ appears only in the terms that represent convection by mean velocity $U(r)$, we are able to identify the parts of the spectrum that hails from the wall, and those that hail from the centreline by comparing the left column of Fig.~\ref{fig:naturalmodes} to that on the right. Since $U(r=1) = 0$, the points of the spectrum that remains unchanged when $\alpha$ is changed should hail from the wall, and those that undergo large shift in the real part must hail from the the centreline. Hence, the wall modes are the least-decaying at very low $W$, and the centre modes become the least-decaying at large $W$. 

Figure~\ref{fig:natmode_lstdcy_omgr_omgi}(a) shows the contours of the decay rate, i.e. the imaginary part of $\omega$ of the least-decaying mode obtained from Eq.~(\ref{eq:natural}) in the plane of $(W,L)$ for $\alpha = 1$. As can be be seen, the decay rate of the least-decaying mode has a weak dependence on $L$ as long as it is higher than the allowed lower bound $(L>\sqrt{3})$. The modes become highly decayed as $W$ is decreased.

Figure~\ref{fig:natmode_lstdcy_omgr_omgi}(b) shows the corresponding phase-speed, $\omega/\alpha$. The observed randomness between two values in a chosen region is due to the fact that there are at least two least-decaying modes that share the same decay rate between two phase speeds, as evident from Fig.~\ref{fig:naturalmodes}. At low $L$ and high $W$ as in Fig.~\ref{fig:naturalmodes}(e)--(f), there may be only one least-decaying mode. However, due to the continuous nature of these modes, it would be difficult to capture the phase-speed due to the round off errors in the decay rate. The message, however, from Figure~\ref{fig:natmode_lstdcy_omgr_omgi}(b) is that at sufficiently large $L$ an increase in W causes a noticeable increase in the phase-speed. This can be tracked to the following facts:

The Figure~\ref{fig:natmode_lstdcy_omgr_omgi}(b) shows that the range of $W$ is $1\sim 31$. At this range, we the modes that are least-decaying originate away from the wall, which is also evident from the discussion on Fig.~\ref{fig:naturalmodes} above. Therefore, at the location of origin of these modes the mean velocity, $U(r)$ is of some finite value. It is known from Ref.~\cite{malik2020viscoelastic} that an increase in $W$ causes an enhancement in shear thinning of the laminar flow causing the $U(r)$ to increase. This increase in $U(r)$ enhances the convection resulting in an increased phase-speed This qualitative dependence of phase speed on $U(r)$ is also evident from the real part of $\omega$ expressed in analytical form later (see Eq.~(\ref{eq:omg_analytic}) together with the definition of $b(r)$). 

The apparent three branches in each panel in Fig.~\ref{fig:naturalmodes} arise from the three roots of the cubic characteristic relation that is obtained from Eq.~(\ref{eq:natural}) as a condition of existence of a non-trivial solution. This relation can be written as
\begin{equation}
\omega^3 + b\omega^2 +\overline{c}\omega+d = 0,
\end{equation}
where $b(r) = -3\alpha U + {\rm i}(WF^3)^{-1}(3F^2G+3F^4+8GW^2r^2)$, 
$\overline{c}(r) = 3[(\alpha U)^2-W^{-2}(F^2+2G)-8Gr^2F^{-2}] -2{\rm i}\alpha U(WF^3)^{-1}[3F^2G+3F^4+8GW^2r^2]$,
$d(r) = \alpha U (FW)^{-2}[3F^4+6F^2G+24W^2Gr^2-(W\alpha U F)^2]+{\rm i}(WF)^{-3}\{3F^4(\alpha U W)^2-3F^4G-F^6 + 3GW^2F^2[(\alpha U)^2-8r^2] + 8GW^4(r\alpha U)^2\}$. The three roots corresponding to the observed branches in the spectrum are
\begin{equation}
\omega_k = -\frac{1}{3}\left (b(r_0) + \xi^kB + \frac{\Delta_0}{\xi^kB}\right ) \ \text{for} \ k \in \{0,1,2\} \ \text{and} \ r_0\in [0,1], \label{eq:omg_analytic}
\end{equation} 
where $\xi = (-1 +{\rm i}\sqrt{3})/2$, $\Delta_0 = b(r_0)^2-3\overline{c}(r_0)$, $B = ([\Delta_1+\sqrt{\Delta_1^2-4\Delta_0^3}]/2)^{1/3}$ with any choice for square- and cubic-roots, and $\Delta_1 = 2b(r_0)^3 -9b(r_0)\overline{c}(r_0) + 27d(r_0)$. Note that for each value of $k$, there is an uncountable set of $\omega$'s. We verified that this expression gives the spectrum obtained earlier using QZ algorithm in Fig.~\ref{fig:naturalmodes} and the contours in Fig.~\ref{fig:natmode_lstdcy_omgr_omgi}.

Since the linear operator in Eq.~(\ref{eq:natural}) does not have any derivatives, the eigenfunctions exhibit a pulse-like behaviour that violates differentiability for these modes with respect to $r$. For example, let us consider an eigenvalue of $\omega = \omega_1(r_0)$, where $r_0 \in [0,1]$ is our choice. Then the solution of the system in Eq.~(\ref{eq:natural}), gives the values of $c_{11,1}'(r_0)$, $c_{22,1}'(r_0)$, $c_{12,1}'(r_0)$ (but, up to a multiplicative constant, since Eq.~(\ref{eq:natural}) is homogeneous). The third index with value $1$ in the suffix suggests that we are considering the family of modes with $k = 1$. When $r \neq r_0$, the eigenfunctions satisfy $c_{11,1}'(r)=c_{22,1}'(r)=c_{12,1}'(r)=0$. Since there is an infinite and continuous choice for $r_0$, we have an infinite set of eigenfunctions.

Despite these eigenfunctions' pulse-like behaviour, small disturbances with arbitrary profiles with respect to $r$, say, $\tilde{c}_{11}(r)$, $\tilde{c}_{22}(r)$, $\tilde{c}_{12}(r)$, can be decomposed as $\tilde{c}_{11}(r) = \sum_{k=1}^3\tilde{a}_k(r)c_{11,k}'(r)$, $\tilde{c}_{22}(r) = \sum_{k=1}^3\tilde{a}_k(r)c_{22,k}'(r)$, and $\tilde{c}_{12}(r) = \sum_{k=1}^3\tilde{a}_k(r)c_{12,k}'(r)$, where the expansion coefficients $\tilde{a}_k$'s, can be found from
\begin{equation}
\begin{pmatrix}
\tilde{a}_1(r)\\
\tilde{a}_2(r)\\
\tilde{a}_3(r)
\end{pmatrix}
=
\begin{pmatrix}
c_{11,1}'(r) \ & c_{11,2}'(r) \ & c_{11,3}'(r)\\
c_{22,1}'(r) \ & c_{22,2}'(r) \ & c_{22,3}'(r)\\
c_{12,1}'(r) \ & c_{12,2}'(r) \ & c_{12,3}'(r)
\end{pmatrix}^{-1}
\begin{pmatrix}
\tilde{c}_{11}(r)\\
\tilde{c}_{22}(r)\\
\tilde{c}_{33}(r)
\end{pmatrix}.
\end{equation}

It should be noted that these modes corresponding to $u'(r) = v'(r) = w'(r) = 0$ are present only for the case of $\beta =1$. For the cases of $\beta \neq 1$, $u'(r) = v'(r) = w'(r) = 0$ in  Eq.~(\ref{gov_eqn_u}) adds a constraint,
\begin{equation}
\boldsymbol{\nabla} \times (\boldsymbol{\nabla}\cdot\boldsymbol{\tau'}) = 0, \label{eq:conserv}
\end{equation}
where $\boldsymbol{\tau'} = (F\textbf{\textit{\textsf{c'}}} + \text{tr}(\textbf{\textit{\textsf{c'}}})G\textbf{\textit{\textsf{C}}})/W$. This set of extra conditions in Eq.~(\ref{eq:conserv}) implies that the force due to the polymer term in Eq.~(\ref{gov_eqn_u}) is conservative. Since the total number of conditions that need to be satisfied in Eq.~(\ref{eq:natural}) and Eq.~(\ref{eq:conserv}) is six while the number of unknowns ($c_{11}'$, $c_{22}'$ and $c_{23}'$) is three, the only possibility is the trivial solutions for these unknowns, and hence for $c_{33}'$, $c_{13}'$ and $c_{23}'$. (Note that $c_{33}' = c_{22}'$, and $c_{13}' = c_{23}' = 0$ for all modes and at all $r$ within the domain.)

\section{Petrov--Galerkin spectral method \label{sec:petrov}}
In the linear system given by Eqs.~(\ref{eq:umom})--(\ref{eq:c23t}), pressure is an unknown. There are three broad strategies for eliminating $p'$ from this system. Among those the {\it weaker formulation} of Petrov--Galerkin method is most suitable for the current viscoelastic model since it exhibits shear-thinning. In order to explain this, let us briefly describe those three methods of eliminating $p'$.  

Let us cast Eqs.~(\ref{eq:umom})--(\ref{eq:wmom}) in the vector form 
\begin{equation}
\boldsymbol{h}\left (\boldsymbol{u'}(r), r\right ) = \boldsymbol{\nabla}p'(r), \label{eq:gradp}
\end{equation}
where each component of $\boldsymbol{h}$ corresponds to all the LHS terms of each of Eqs.~(\ref{eq:umom})--(\ref{eq:wmom}) after bringing the RHS terms to LHS except those involving $p'$. In the LHS of Eq.~(\ref{eq:gradp}), we have suppressed the dependence of $\boldsymbol{h}$ on $\textbf{\textit{\textsf{c'}}}$ and other parameters for brevity.

In the first method of eliminating $p'$ from Eq.~(\ref{eq:gradp}), one makes use of the fact that pressure is an example of a potential function, i.e., Eq.~(\ref{eq:gradp}) is replaced by $\boldsymbol{\nabla}\times \boldsymbol{h} = 0$, a result that is due to the identity $\boldsymbol{\nabla}\times \boldsymbol{\nabla} p' = 0$. This strategy has been widely applied for Newtonian pipe flows (see for example~\cite{burridge1969comments,schmid1994optimal,malik2019linear}), and viscoelastic pipe flows under the Oldroyd-B model~\cite{garg2018viscoelastic}. 

In the second method, known as the method of influence matrix~\cite{priymak1998accurate}, the $p'$ in Eq.~(\ref{eq:gradp}) is handled in conjunction with the divergence of Eq.~(\ref{eq:gradp}), i.e., $\boldsymbol{\nabla\cdot h} = \nabla^2p'$, where on the LHS, the continuity condition $\boldsymbol{\nabla \cdot u'}=0$ is used to eliminate the unsteady and viscous terms. For further description of solving these two equations, see Priymak \& Miyazaki~\cite{priymak1998accurate}.

The third strategy is known as Petrov--Galerkin method, where Eq.~(\ref{eq:gradp}) is solved in its weak form. In this method, the solutions are expanded in a known set of $M+1$ vector-valued functions, $\boldsymbol{u_j'}(r)$ $(j = 0\cdots M)$, which form {\it trial basis}. Eq.~(\ref{eq:gradp}) is then projected on another known set of functions, $\boldsymbol{\hat{u}_j'}(r)$ $(j = 0\cdots M)$, named as {\it test functions}. These sets of functions, $\{\boldsymbol{u_j'}(r)\}$ and $\{\boldsymbol{\hat{u}_j'}(r)\}$, which will be specified later, need to be different due to the presence of odd order derivatives in the convective and polymer terms in Eqs.~(\ref{eq:umom})--(\ref{eq:c23t}), hence the term, {\it Petrov--Galerkin}. These $\boldsymbol{u_j'}(r)$'s and $\boldsymbol{\hat{u}_j'}(r)$'s are chosen as solenoidal functions, i.e., $\boldsymbol{\nabla\cdot u_j'}= \boldsymbol{\nabla\cdot \hat{u}_j'}=0$. The result is
\begin{equation}
\int_0^1r \diff r  [\boldsymbol{\hat{u}_i'}(r)]^{\rm H}\sum_{j = 0}^M a_j\boldsymbol{h}\left (\boldsymbol{u'_j}(r), r\right ) = 0 \ \text{for} \ \text{each}\ i\in (0\cdots M), \label{eq:mickey_projection}
\end{equation}
where $a_j$'s are expansion coefficients, and the superscript, `H', refers to the Hermitian conjugate. The RHS of Eq.~(\ref{eq:mickey_projection}) is zero, since the RHS of Eq.~(\ref{eq:gradp}) vanishes under these operations, which is in turn due to 
\begin{align}
\int_0^1r \diff r \left ([\boldsymbol{\hat{u}_i'}(r)]^{\rm H}\boldsymbol{\nabla}p'(r) \right ) =
&- \int_0^1r \diff r \left ([\boldsymbol{\nabla\cdot\hat{u}_i^{'*}}(r)]p'(r) \right ) = 0, \label{eq:p_vanish}
\end{align}
where we have used integration by parts, and conditions of no-slip, no-penetration and continuity. The superscript, `*' refers to complex-conjugate.

Now the system~(\ref{eq:mickey_projection}) is free from pressure fluctuations, $p'$. This way of eliminating $p'$ is advantageous for the FENE-P model since it does not require an increase of the order of differentiations, unlike in the first two methods described above. Under the FENE-P model, the restoring tendency, $F(r)$ (i.e., the Peterlin function), is dependent on $C_{11}(r)$, $C_{22}(r)$ and $C_{33}(r)$. Further, $G_1(r)$ and $G_2(r)$ in Eqs.~(\ref{eq:umom})--(\ref{eq:vmom}) are also functions of the mean conformation tensor and the function $G(r)$, which itself is a function of $F(r)$. The process of subjecting Eq.~(\ref{eq:gradp}) to the operation of the curl as required in the first method, or to that of divergence as needed in the second method, increases the complexity by producing numerous terms by chain- and product-rules for differentiation. Hence, we choose the method of Petrov--Galerkin.

Now we describe this Petrov-Galerkin method adopted here in more detail. Let us write the nine Eqs.~(\ref{eq:umom})--(\ref{eq:c23t}) in the form,
\begin{equation}
\boldsymbol{g}\left (\boldsymbol{\phi'}(r), r\right ) = -{\rm i}\omega \boldsymbol{\phi'}(r), \label{eq:gov_phi}
\end{equation}
where $\boldsymbol{\phi'}(r) = \{u', v', w', c_{11}', c_{22}', c_{33}', c_{12}', c_{13}', c_{23}'\}^{\rm T}$, and $\boldsymbol{g}(\boldsymbol{\phi'}(r), r)$ is a 9-tuple array-valued function with each element corresponding to the RHS of each of Eqs.~(\ref{eq:umom})--(\ref{eq:c23t}) in the same order, after bringing the LHS terms to the RHS except for the terms involving $\omega$.  Further, in the first three elements of the array $\boldsymbol{g}(\boldsymbol{\phi'}(r), r)$, the terms involving $p'$ are neglected, since they do not contribute upon projection on the solenoidal test functions as described above in Eq.~(\ref{eq:p_vanish}) (again, as in the case of $\boldsymbol{h}$ above, we have suppressed $\boldsymbol{g}$'s dependence on parameters, $\{\alpha, n, \Rey, W, L, \beta\}$ and other mean-flow variables for brevity). 

As before, let us consider the trial functions $\boldsymbol{\phi_j'}(r)$ with $j \in (0,\cdots M)$ such that 
\begin{equation}
\boldsymbol{\phi'}(r)=\sum_{j = 0}^M a_j\boldsymbol{\phi_j'}(r), \label{eq:trial_expansion}
\end{equation}
and the test functions with notation $\boldsymbol{\hat{\phi}_j'}(r)$ with $j \in (0,\cdots M)$. These functions $\boldsymbol{\phi_j'}(r)$'s and $\boldsymbol{\hat{\phi}_j'}(r)$'s are specified later in this section.  The weak form of Eq.~(\ref{eq:gov_phi}) that corresponds to Eq.~(\ref{eq:mickey_projection}) is given by
\begin{equation}
\int_0^1r \diff r  [\boldsymbol{\hat{\phi}_i'}(r)]^{\rm H}\sum_{j = 0}^Ma_j\boldsymbol{g}\left (\boldsymbol{\phi'_j}(r), r\right ) = -{\rm i}\omega\int_0^1r \diff r  [\boldsymbol{\hat{\phi}_i'}(r)]^{\rm H}\sum_{j = 0}^Ma_j\boldsymbol{\phi'_j}(r), \label{eq:projection}
\end{equation}
for each $i\in (0\cdots M)$. The functions $\boldsymbol{\phi_j'}(r)$'s and $\boldsymbol{\hat{\phi}_j'}(r)$'s are members of sets mentioned below.
\begin{align}
\boldsymbol{\phi_j'}(r) &\in \{\boldsymbol{\phi_l^{(m)}}(r), \ \ m\in(1,\cdots,8), \ \ l\in(0,\cdots,N) \} \label{eq:trial_set} \\
\boldsymbol{\hat{\phi}_j'}(r) &\in \{\boldsymbol{\hat{\phi}_l^{(m)}}(r), \ \ m\in(1,\cdots,8), \ \ l\in(0,\cdots,N) \},
\end{align}
Eq.~(\ref{eq:trial_expansion}) and Eq.~(\ref{eq:trial_set}) imply that $M+1 =8(N+1)$. The functions $\boldsymbol{\phi_l^{(m)}}(r)$ are chosen such that they follow the ansatzes in Eqs.~(\ref{eq:vel_behaviour})--(\ref{eq:cij_behaviour}) and the continuity condition. The functions $\boldsymbol{\hat{\phi}_l^{(m)}}(r)$ need not follow these ansatz, but need to be divergence free. These $\boldsymbol{\hat{\phi}_l^{(m)}}(r)$'s are chosen such that the integrands in Eq.~(\ref{eq:projection}) are even, since doing so makes these test functions different from the trial functions, a requirement for this Petrov--Galerkin method to capture the non-self-adjointness of the system in Eqs.~(\ref{eq:umom})--(\ref{eq:c23t}) (the even nature of the integrands also helps in choosing a numerical scheme for radial discretization that avoids clustering of points at the pipe centreline as found by Meseguer and Trefethen~\cite{meseguer2000spectral}). This implies that every element of the 9-tuple array $\boldsymbol{\hat{\phi}_j^{(m)}}(r)$ should not have the same parity as that of $\boldsymbol{\phi_j^{(m)}}(r)$ (note that every element of the 9-tuple array $\boldsymbol{g}\left (\boldsymbol{\phi'_j}(r), r\right )$ is in same parity as that of $\boldsymbol{\phi_j^{(m)}}(r)$ by Eq.~(\ref{eq:gov_phi}) and by the findings in Subsect.~\ref{sec:heur}).

The first three elements of 9-tuple arrays $\boldsymbol{\phi_j^{(m)}}(r)$ and $\boldsymbol{\hat{\phi}_j^{(m)}}(r)$ for a chosen $j$ and $m$ constitute the velocity perturbations, $\boldsymbol{u_j^{(m)}}(r)$ and $\boldsymbol{\hat{u}_j^{(m)}}(r)$ that follow the solenoidal conditions, $\boldsymbol{\nabla\cdot u_j^{(m)}}= \boldsymbol{\nabla\cdot \hat{u}_j^{(m)}}=0$, and the boundary conditions, $\boldsymbol{u_j^{(m)}}(r) = \boldsymbol{\hat{u}_j^{(m)}}(r) = 0$ for $r \in (0,1)$. We choose these three components identical to Meseguer and Trefethen~\cite{meseguer2000spectral,meseguer2003linearized} for each $j$ and $m$ (except for some minor changes in the test functions detailed later). These chosen functions are given below for each cases:
\begin{align}
&
\left (
\begin{array}{l}
n=0\\ 
\alpha=0
\end{array}
\right )
: \left \{
\begin{array}{ll}
\boldsymbol{\phi_l^{(m)}}(r) = \{&\psi_l\delta_{1m}, 0, r\psi_l\delta_{2m},r^2T_{2l}\delta_{3m}, r^2T_{2l}\delta_{4m}, r^2T_{2l}\delta_{5m}, \\
&rT_{2l}\delta_{6m}, r^3T_{2l}\delta_{7m}, r^2T_{2l}\delta_{8m}\}^{\rm T}\\
\boldsymbol{\hat{\phi}_l^{(m)}}(r) = \{& r\psi_l\delta_{1m}, 0, \psi_l\delta_{2m},rT_{2l}\delta_{3m}, rT_{2l}\delta_{4m}, rT_{2l}\delta_{5m}, \\
&T_{2l}\delta_{6m}, T_{2l}\delta_{7m}, rT_{2l}\delta_{8m}\}^{\rm T}
\end{array}
\right . \nonumber \\
&
\left (
\begin{array}{l}
n=0\\ 
\alpha\neq0
\end{array}
\right )
: \left \{
\begin{array}{ll}
\boldsymbol{\phi_l^{(m)}}(r) =& \{D_+[r\varphi_l]\delta_{1m}, -{\rm i}\alpha r\varphi_l\delta_{1m}, r\psi_l\delta_{2m},\\ 
&T_{2l}\delta_{3m}, T_{2l}\delta_{4m}, T_{2l}\delta_{5m}, rT_{2l}\delta_{6m}, rT_{2l}\delta_{7m}, r^2T_{2l}\delta_{8m}\}^{\rm T} \\
\boldsymbol{\hat{\phi}_l^{(m)}}(r) = & \{(D_+[r^2\varphi_l]+r^3\psi_l)\delta_{1m}, -{\rm i}\alpha r^2\varphi_l\delta_{1m}, \psi_l\delta_{2m},\\ 
&rT_{2l}\delta_{3m}, rT_{2l}\delta_{4m}, rT_{2l}\delta_{5m},T_{2l}\delta_{6m}, T_{2l}\delta_{7m}, rT_{2l}\delta_{8m}\}^{\rm T}
\end{array}
\right . \nonumber \\
&(|n|=1): \left \{
\begin{array}{ll}
\boldsymbol{\phi_l^{(m)}}(r) =& \{nr\psi_l\delta_{1m}, -{\rm i}n \varphi_l\delta_{2m}, -\alpha r^2\psi_l\delta_{1m}+D[r\varphi_l]\delta_{2m},\\ 
&rT_{2l}\delta_{3m}, rT_{2l}\delta_{4m}, rT_{2l}\delta_{5m}, T_{2l}\delta_{6m}, T_{2l}\delta_{7m}, rT_{2l}\delta_{8m}\}^{\rm T} \\
\boldsymbol{\hat{\phi}_l^{(m)}}(r) = & \{n\psi_l\delta_{1m}, -{\rm i}n r\varphi_l\delta_{2m}, -\alpha r\psi_l\delta_{1m}+(D[r^2\varphi_l]+r^3\psi_l)\delta_{2m},\\
&T_{2l}\delta_{3m}, T_{2l}\delta_{4m}, T_{2l}\delta_{5m}, rT_{2l}\delta_{6m}, rT_{2l}\delta_{7m}, T_{2l}\delta_{8m}\}^{\rm T} 
\end{array}
\right . \nonumber \\
&(|n|\geq2): \left \{
\begin{array}{ll}
\boldsymbol{\phi_l^{(m)}}(r) = &\{nr^{|n|}\psi_l\delta_{1m}, -{\rm i}nr^{|n|-1} \varphi_l\delta_{2m},\\
& -\alpha r^{|n|+1}\psi_l\delta_{1m}\!+\!D[r^{|n|}\varphi_l]\delta_{2m},\\ 
&r^{|n|}T_{2l}\delta_{3m}, r^{|n|-2}T_{2l}\delta_{4m}, r^{|n|-2}T_{2l}\delta_{5m}, \\
&r^{|n|-1}T_{2l}\delta_{6m}, r^{|n|-1}T_{2l}\delta_{7m}, r^{|n|-2}T_{2l}\delta_{8m}\}^{\rm T} \\
\boldsymbol{\hat{\phi}_l^{(m)}}(r) = & \{nr^{\gamma_1}\psi_l\delta_{1m}, -{\rm i}nr^{\gamma_2} \varphi_l\delta_{2m},  \\
&-\alpha r^{\gamma_1+1} \psi_l\delta_{1m} \!+\! (D[r^{\gamma_2+1}\varphi_l] + r^{\gamma_2+2}\psi_l) \delta_{2m},\\ 
&r^{\gamma_1}T_{2l}\delta_{3m}, r^{\gamma_1}T_{2l}\delta_{4m}, r^{\gamma_1}T_{2l}\delta_{5m}, \\
&r^{\gamma_2}T_{2l}\delta_{6m}, r^{\gamma_2}T_{2l}\delta_{7m}, r^{\gamma_1}T_{2l}\delta_{8m}\}^{\rm T} 
\end{array}
\right . \nonumber 
\end{align}
where $\delta_{ij}$ is Kronecker delta, $T_{2l}(r)$ is the Chebyshev polynomial of order $2l$,  $\psi_l(r) = (1-r^2)T_{2l}(r)$, $\varphi_l(r) = (1-r^2)^2T_{2l}(r)$, $D_+ = (r+D)$, $(\gamma_1,\gamma_2) = (1,0)$ for even $n$, and $(\gamma_1,\gamma_2) = (0,1)$ when $n$ is odd. The velocity components of the test functions differ slightly from that of Meseguer \& Trefethen~\cite{meseguer2000spectral} for the following reasons. We have dropped the Chebyshev weights $(1-r^2)^{-1/2}$, since we confirmed after testing that these get cancelled out owing to the fact that the integral acts on both sides of the equality symbol in Eq.~(\ref{eq:projection}) and that these weights would be on both sides. Second, the $\hat{u}$ and the $\hat{w}$ components of the test functions for the cases of odd $n$ and $m = 1$ (i.e., the first and third elements of $\boldsymbol{\hat{\phi}_l^{(1)}}$) are identical to Ref.~\cite{meseguer2000spectral} (i.e., in their notation, the third and second components of $\tilde{v}_m^{(2)}$ of Eq.~2.54 in Ref.~\cite{meseguer2000spectral}, respectively), but after factoring out a $r^2$. This does not affect the requirement that the test functions need to be solenoidal with opposite parity to that of trial functions, but can improve the accuracy since the higher powers of $r$ can lead to round-off errors close to the centreline. 

Eq.~(\ref{eq:projection}) can be written as
\begin{equation}
\mathcal{A} \boldsymbol{a} = -{\rm i}\omega \mathcal{B} \boldsymbol{a}, \label{eq:eigsys}
\end{equation}
where $\mathcal{A}$ and $\mathcal{B}$ are matrices with elements, $\mathcal{A}_{ij} = \int_0^1r \diff r  [\boldsymbol{\hat{\phi}_i'}(r)]^{\rm H}\boldsymbol{g}\left (\boldsymbol{\phi'_j}(r), r\right )$, and $\mathcal{B}_{ij} = \int_0^1r \diff r  [\boldsymbol{\hat{\phi}_i'}(r)]^{\rm H}\boldsymbol{\phi'_j}(r)$. For radial discretization scheme, we follow Meseguer \& Trefethen~\cite{meseguer2000spectral}, where the Gauss Lobatto grid, and a clever exploitation of parities has been adopted. The eigensystem in Eq.~(\ref{eq:eigsys}) is solved using the QZ algorithm implemented in MATLAB (version R2020a). We use $N =100$ for the computations in this paper. For the radial discretization, we use $2(N+|n|)+9$ number of Gauss-Lobatto collocation points following Ref.~\cite{meseguer2000spectral}.

Since this is the first work on the linear perturbations to a shear-thinning viscoelastic pipe flow, we are able to validate our code only for the case of $\beta = 1$ against the Newtonian flow results of Meseguer and Trefethen~\cite{meseguer2000spectral}. As explained in Sect.~\ref{sec:natural}, for $\beta = 1$, the Newtonian spectrum is a subset of the whole spectrum from Eq.~(\ref{eq:eigsys}), with the rest being the natural spectrum of polymers corresponding to the trivial solutions for velocity perturbations. 
\begin{table*}\centering
\caption{{Comparison of least-decaying modes for $\beta = 1$, $W = 0.1$, $L = 5$, and $\Rey = 3000$ with that of Newtonian flow}\label{tb:compare_meseg}}.
{\footnotesize
\begin{tabular}{|cc|ll|ll|}
\hline
\multicolumn{2}{|c|}{Wave-} & \multicolumn{2}{c|}{From the present} & \multicolumn{2}{c|}{Newtonian result}\\
\multicolumn{2}{|c|}{numbers} & \multicolumn{2}{c|}{FENE-P code} & \multicolumn{2}{c|}{(Meseguer \& Trefethen, 2000)}\\
\hline
\multicolumn{1}{|c}{$n$}& \multicolumn{1}{c}{$\alpha$} & \multicolumn{1}{|c}{$\omega_r$} & \multicolumn{1}{c|}{$\omega_i$} & \multicolumn{1}{|c}{$\omega_r$} & \multicolumn{1}{c|}{$\omega_i$} \\
\hline
$0$ & $0$ & \multicolumn{1}{c}{$0$} & $-0.00192772865$ & \multicolumn{1}{c}{$0$} & $-0.00192772865$\\
$0$ & $1$ & $-0.94836022205$ & $-0.05197311128$ &$-0.94836022205$  & $-0.05197311128$\\
$1$ & $1$ & $-0.91146556762$ & $-0.04127564467$ &$-0.91146556762$  & $-0.04127564469$\\
$2$ & $1$ & $-0.88829765890$ & $-0.06028568944$ &$-0.88829765875$  & $-0.06028568956$\\
$3$ & $1$ & $-0.86436392102$ & $-0.08325397697$ &$-0.86436392104$  & $-0.08325397694$\\
\hline
\end{tabular}
}
\end{table*}
We use this fact to validate the present code against the least decaying modes of the Newtonian case in Table~\ref{tb:compare_meseg}. The natural spectrum of the polymers discussed in Sect.~\ref{sec:natural} that have $u'(r) = v(r) = w'(r) = 0$ can also be expected to appear from this Petrov--Galerkin code, but they will be highly contaminated owing to the fact that these modes' eigenfunctions are non-zero only at a point in the radial domain, and that the Chebyshev spectral expansion of such non-differentiable functions will be poorly represented. Therefore, we choose $W$$(=0.1)$ such that these modes (which exist only for this case of $\beta = 1$ as explained in Sect.~\ref{sec:natural}) are pushed to the most-decayed regions of the spectrum. 

As can be seen from Table~\ref{tb:compare_meseg}, the least-decaying modes for the case of $\beta=1$ are in good agreement until the 9-th decimal place with the Newtonian result of Meseguer and Trefethen~\cite{meseguer2000spectral} for several combinations of $n$ and $\alpha$. The symbols $\omega_r$ and $\omega_i$ are defined by $\omega = \omega_r+{\rm i}\omega_i$. (Note that Meseguer and Trefethen~\cite{meseguer2000spectral} have tabulated the values of $-{\rm i}\omega$ from which the values of $\omega$ are obtained. We have flipped the sign of $\omega_r$ of Meseguer and Trefethen~\cite{meseguer2000spectral} in Table~\ref{tb:compare_meseg} to conform to the present convention that the mean flow is directed along the axial unit vector, which is opposite to that of Ref.~\cite{meseguer2000spectral}.) 

\begin{table*}\centering
\caption{{Least-decaying modes for $W = 0.1$, $L = 5$, $\Rey = 3000$ and $\beta = 0.8$ and $0.9$.}\label{tb:beta_comp}}
{\footnotesize
\begin{tabular}{|cc|ll|ll|}
\hline
\multicolumn{2}{|c|}{$n,\alpha$} & \multicolumn{2}{c|}{$\beta = 0.8$} & \multicolumn{2}{c|}{$\beta=0.9$}\\
\hline
\multicolumn{1}{|c}{$n$}& \multicolumn{1}{c}{$\alpha$} & \multicolumn{1}{|c}{$\omega_r$} & \multicolumn{1}{c|}{$\omega_i$} & \multicolumn{1}{|c}{$\omega_r$} & \multicolumn{1}{c|}{$\omega_i$} \\
\hline
$0$ & $0$ & \multicolumn{1}{c}{$0$} & $-0.00154220380$ & \multicolumn{1}{c}{$0$} & $-0.00173496620$\\
$0$ & $1$ & $0.95415302704$ & $-0.04648587912$ &$0.95118028015$  & $-0.04930543538$\\
$1$ & $1$ & $0.92039736589$ & $-0.03730175440$ &$0.91582064235$  & $-0.03934027033$\\
$2$ & $1$ & $0.89936047235$ & $-0.05448104342$ &$0.89369207726$  & $-0.05745655055$\\
$3$ & $1$ & $0.87766487357$ & $-0.07525941202$ &$0.87084999306$  & $-0.07935670524$\\
\hline
\end{tabular}
}
\end{table*}
Table~\ref{tb:beta_comp} shows the least-decaying modes for $\beta = 0.8$ and $0.9$, and demonstrates that an increase in $\beta$ decreases the phase-speed of these modes for various combinations of $n$ and $\alpha$, but makes the perturbations more stable. Since increasing $\beta$ weighs the viscous contribution from the solvent to the total stress more than that of the elastic contribution from the polymers, this trend indicated by the Table~\ref{tb:beta_comp} implies that the addition of the polymers would destabilize the Tollmien-Schlichting waves of the solvent flow. 

\begin{table*}\centering
\caption{{ Modes with highest growth rate for $\beta = 0.9$, $L = 5.0$ and $\Rey = 3000$.}\label{tb:W_comp}}
{\footnotesize
\begin{tabular}{|cc|ll|ll|}
\hline
\multicolumn{2}{|c|}{$n,\alpha$} & \multicolumn{2}{c|}{$W = 1$} & \multicolumn{2}{c|}{$W=50$}\\
\hline
\multicolumn{1}{|c}{$n$}& \multicolumn{1}{c}{$\alpha$} & \multicolumn{1}{|c}{$\omega_r$} & \multicolumn{1}{c|}{$\omega_i$} & \multicolumn{1}{|c}{$\omega_r$} & \multicolumn{1}{c|}{$\omega_i$} \\
\hline
$0$ & $0$ & \multicolumn{1}{c}{$0$} & $1.3933687519$ & \multicolumn{1}{c}{$0$} & $2.75476338086$\\
$0$ & $1$ & $0.00013169023$ & $1.3928992117$ &$0.00147249809$  & $2.75397719258$\\
$1$ & $1$ & $0.00012717590$ & $1.3928997764$ &$0.00144798717$  & $2.75387148923$\\
$2$ & $1$ & $0.00012631528$ & $1.3928994817$ &$0.00144867521$  & $2.75384296713$\\
$3$ & $1$ & $0.00012616687$ & $1.3928991079$ &$0.00145074024$  & $2.75382987298$\\
\hline
\end{tabular}
}
\end{table*}
In Table~\ref{tb:W_comp}, the fastest-growing modes are shown for two values of $W$, i.e., $1$ and $50$, with other parameters fixed at $\beta = 0.9$, $L = 5$ and $\Rey = 3000$. Upon comparing the least-decaying/fastest-growing modes for the case of $W=1$ in this table with the same in Table~\ref{tb:beta_comp} for $\beta = 0.9$, we see that a strong instability with $\mathcal{O}(1)$ growth rate occurs even at such low value as $W = 1$. We note that this transition occured in the interval $0.4<W<0.5$ (not shown). A detailed study of the cause for this instability through an energy budget analysis would be required for an understanding the complete picture. 

In addition, the comparison of the fastest-growing modes in Table~\ref{tb:W_comp} for various combinations of the wavenumbers suggest that the growth rates and phase speeds are largely insensitive to the changes in wavenumbers. 

Comparing the fastest growing modes shown in Table~\ref{tb:W_comp} for the two different values of $W$, one learns that an increase in $W$ causes the flow to be more unstable. This can be explained from the perspective of shear-thinning. As mentioned in Sect.~\ref{sec:natural}, an increase in $W$ increases the flow rate due to shear-thinning. We should take into consideration that the $\Rey$ is defined here such that the pressure gradient is kept constant when $W$ is changed and is indifferent to the increase in flow rate due to shear thinning during the course of change in $W$.  However, the $\Rey$ defined based on the centreline velocity (say, $\Rey_c$) of the current non-Newtonian flow will increase with $W$. It's known that an increase in $\Rey_c$ would enhance the growth rate due to the availability of more energy, which in turn is due to a decrease in the viscous dissipation. However, a detailed study is warranted as a future goal to study this using energy budget analysis since there are other changes in the mean-flow variables, such as $U_r$.

\begin{figure}\centering
	\includegraphics[width=0.8\textwidth]{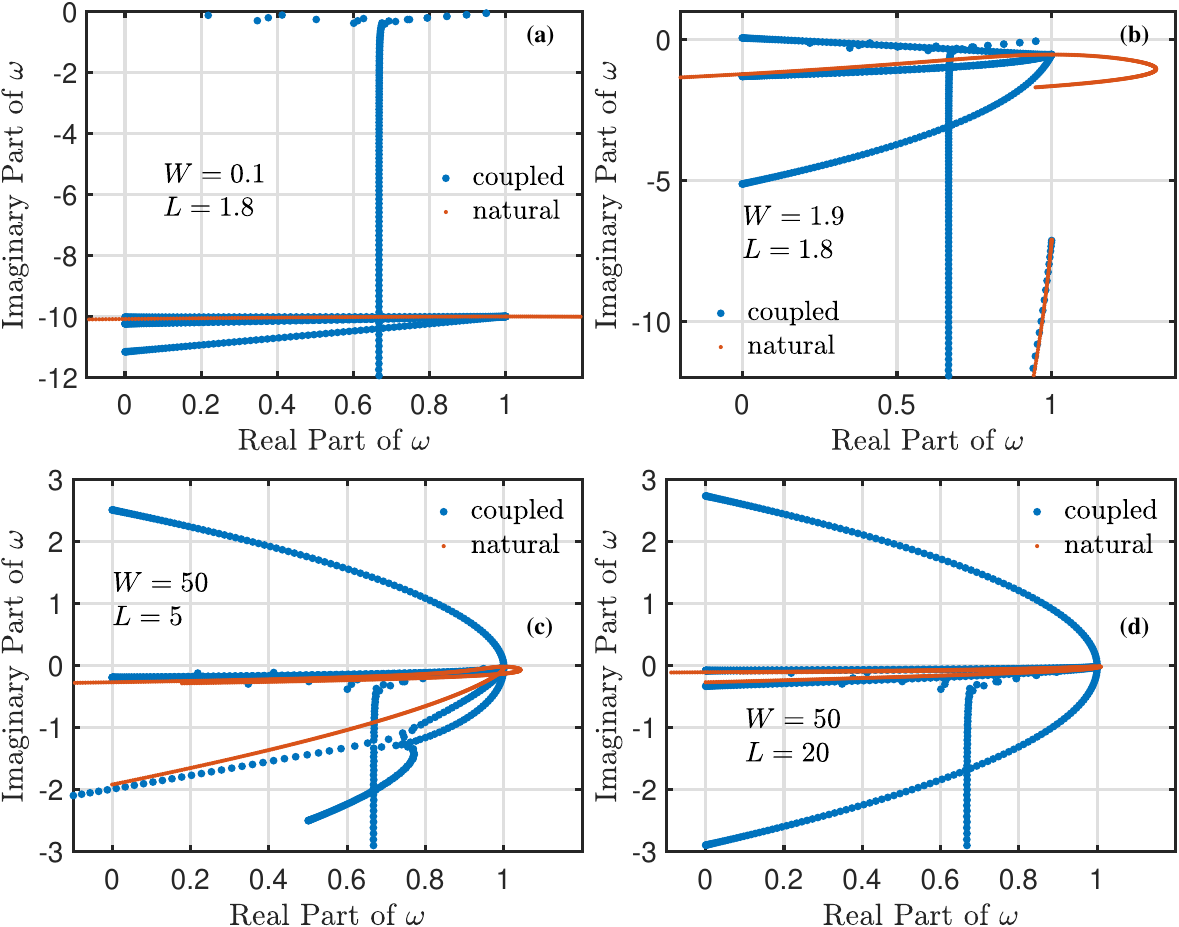}
	\caption{The spectrum (blue spots) for $n = 1$, $\alpha = 1.0$, $\Rey = 3000$, $\beta = 0.99999$: (a) $W = 0.1$ and $L = 1.8$; (b) $W=1.9$ and $L=1.8$; (c) $W=50$ and $L=5.0$; (d) $W=50$ and $L=20$. The brown line is the natural spectrum of polymers from Eq.~(\ref{eq:natural}) for $\beta = 1$. \label{fig:spectrum_n1al1}}
\end{figure}
Figure~\ref{fig:spectrum_n1al1} shows the least-decaying/fastest-growing portions of the spectrum for $\beta = 0.99999$ obtained from the Petrov--Galerkin code and superimposed with the natural spectrum obtained from Eq.~(\ref{eq:natural}) for $\beta = 1$. The aim of showing this figure is in two fold: First, even with such a tiny contribution from polymers to the total viscosity, the instability arises. Second, the spectrum of the coupled system given by Eq.~(\ref{eq:umom})--(\ref{eq:c23t}) still have the branches that exhibit some similarity to the natural spectrum obtained from Eq.~(\ref{eq:natural}). However, since the solution $\boldsymbol{u'}(r) = 0$ is not possible for $\beta\neq 0$, as explained in Sect.~\ref{sec:natural}, these modes contain non zero velocity components. 

Figures~\ref{fig:spectrum_n1al1}(a) and~\ref{fig:spectrum_n1al1}(b) show the spectrum for $L=1.8$, a value that is close the allowed minimum value of $\sqrt{3}$. The values of $W$ in these panels have been chosen such that the spectrum is globally stable in Fig.~\ref{fig:spectrum_n1al1}(a) with $W=0.1$ and is unstable in Fig.~\ref{fig:spectrum_n1al1}(b) at the lowest possible value of $W$($=1.9$ at this value of $\beta$). As it can be noted from these panels, there are branches of the coupled spectrum from Eq.~(\ref{eq:eigsys}) that resembles the natural spectrum from Eq.~(\ref{eq:natural}). The coupled spectrum also have new branches that are different from the corresponding equivalence of natural modes. The observed unstable modes hail from these new branches of the spectrum. In Fig.~\ref{fig:spectrum_n1al1}(b) the instability occurs at the region where the phase-speed is closer to zero. This is in contrast to the observations by Khalid {\it et al.}~\cite{khalid2021centre}, in the case of plane Poiseuille flow, that the instabilities are due to modes that become unstable with phase-speed closer to one.

Figures~\ref{fig:spectrum_n1al1}(c) and~\ref{fig:spectrum_n1al1}(d) show the spectra for higher values of $W$ and $L$. When comparing the two panels, we see that an increase in $L$ destabilizes further the already unstable modes. This phenomenon can again be interpreted using shear-thinning. It is well known that the stretching of the polymers in the axial direction causes the shear-thinning (for a more recent discussion of this phenomenon, see~\cite{malik2020viscoelastic}). Therefore, a higher value of $L$ allow more stretching and hence a high flow rate and a higher value of $\Rey_c$ (i.e., the Reynolds number defined based on centreline velocity). This argument is similar to the reason discussed above for the observed higher growth-rates for higher values of $W$ in Table.~\ref{tb:W_comp}.

\section{Conclusion and outlook}

In this paper, we have shown that the centreline behaviour of the components of  $\textbf{\textit{\textsf{c'}}}(r)$ exhibits odd or even parities depending on $n$, similarly to that of the velocity field. The leading order behaviours of the components of $\textbf{\textit{\textsf{c'}}}(r)$ are forced upon by that of the velocity field, $\boldsymbol{u'}(r)$. The elucidation of the leading order behaviour and the parities plays a crucial role for regularizing the governing equations and carrying out the stability analysis in a robust manner. 

We found that the near-centre behaviour of the the pressure and velocity fluctuations are identical to their Newtonian counterparts. The reason for this can be tracked as follows. The leading order power law behaviour and parities of $(v', w')$ $\text{---}$ a vector perpendicular to axial direction, and $p'$ $\text{---}$ a scalar, are independent of physical problems~\cite{lewis1990physical}. However, such behaviours of the $u'$ is determined by the governing equations of the problem at hand, i.e., the continuity equation in the present case. Since this equation is in both Newtonian and non-Newtonian cases, the Newtonian behaviours of for these variables were preserved in the non-Newtonian case.

As observed in the Newtonian case, the use of the derived symmetries do not only aid in avoiding spurious modes, but also help improving the accuracy of the solutions and eigenvalues, especially for large azimuthal wavenumbers as demonstrated in~\cite{malik2019linear}. 

The modes with $|n|=1$ are important to flow instability in pipe geometry. The Newtonian limit of this flow exhibits least-decaying modes for $|n| = 1$, after setting aside the case of Stokes modes with $(n=0; \alpha=0)$, which are immune to transient growth~\cite{malik2019linear}.  For $|n| = 1$, we observed that $c_{12}'$ and $c_{13}'$ do not vanish at the centreline, but hold $(c_{12}' +n\mi c_{13}')_{r=0} = 0$ as a consequence of the uniformity of the traction vector in the azimuthal direction close to the centreline. This behaviour is similar to that of the velocity components $v'$ and $w'$ that hold $(v' + n\mi w')_{r=0}=0$. Nonetheless, with the leading-order behaviours for the other components obtained in this work, this condition, $(c_{12}' +n\mi c_{13}')_{r=0} = 0$ is automatically satisfied by Eq.~\eqref{eq:c12t} and Eq.~\eqref{eq:c13t}. Lastly, since the symmetries of the mean profiles of the FENE-P model are also applicable to the Oldroyd-B model, the findings are also applicable there.

{
The continuous natural spectrum that has been uncovered here for $\beta = 1$ has trivial velocity perturbations, and perfect pulse-like perturbations to the conformation tensor that are non-zero only at a radial location. Thus, the phase-speed and growth rates are stratified in the radial direction. These modes with trivial velocity components have been shown to exist only for this case of $\beta=1$, since $\beta\neq 0$ imposes additional conditions rendering the number of equations more than the number of independent unknowns. 

The natural continuous modes are determined by the 2D components of the conformation tensor: $c_{11}'$, $c_{22}'$ and $c_{12}'$. For $\beta \neq 1$, the unstable modes hail from branches that are connected to the discrete manifestations of the natural modes as discussed in Sect.~\ref{sec:petrov}. These two facts motivate further studies their relevance to the elasto-inertial turbulence(EIT), since it has been found to be of 2D~\cite{sid2018two} in nature  and that a subsequent 2D DNS simulation has been able to reproduce EIT~\cite{gillissen2019two}. 

The Petrov--Galerkin spectral method has advantages for FENE-P model over the solenoidal formulation with two unknowns, or over the method of influences matrix as explained in Sect.~\ref{sec:petrov}, and it comes handy to implement the ansatzes derived. The preliminary results show that the pipe flow under FENE-P model is also unstable as predicted for Oldroyd-B models~\cite{garg2018viscoelastic,chaudhary2021linear}. However, the flow exhibits stability for very low values of Weissenberg number.

Our future direction is to carry out a detailed study on the instabilities in the parameters space of $\{n, \alpha, \Rey, \beta, W, L\}$ and compare against the non-shear-thinning Oldroyd-B model results of Chaudhary {\it et al.}~\cite{chaudhary2021linear}. Since the absolute value of $U_r(r)$ in the present shear thinning model is larger than that for a Newtonian flow (as shown in the mean-flow plots in Ref.~\cite{malik2020viscoelastic}), the lift-up effect and Orr mechanism are expected to be more pronounced in the present case, hence a larger transient growth. Therefore, a detailed non-modal analysis also would be of interest to the research community.
}

\begin{acknowledgements}
The authors gratefully acknowledge the fruitful feedback from Professor Rich R Kerswell, DAMPT, University of Cambridge on an earlier draft version of this manuscript.
\end{acknowledgements}

\section*{Declarations}
{\bf Funding:} Not applicable; {\bf Conflict of interest:} No conflict of interests; {\bf Availability of data and material:} All data required are included in this article; {\bf Code Availability:} The codes for Sect.~4--6 available on request. The Petrov--Galerkin code of Sect.~7 will be published in Github in due course after the publication of a manuscript under preparation.

\appendix
{
\section{Derivation of analytic behaviour for $|n|\geq 2$}\label{appA}
To address the regular singularity in  Eqs.~\eqref{eq:umom}--\eqref{eq:wmom}, let $p' = r^s\rchi_0(r)$ and $u' = r^m \rchi_1(r)$ where $m\geq0$ and $s\geq0$ are integers to be determined. The $\rchi_j(r)$'s are general analytic functions with Taylor expansion $\rchi_j(r) = \sum_{k = 0}^{\infty}r^k\rchi_{j,k}$. These $\rchi_j(r)$'s are shown to be even with respect to $r$ in what follows. 

With radial vorticity $\eta' \equiv r^{-1}nu'-\alpha w'$, we have $u' = \mi r d^{-1}(\alpha v' + \alpha r Dv' -n\eta')$ and $w' = \mi d^{-1}(n v' + n r Dv'+\alpha^2r\eta')$~\cite{schmid2001stability}. An analysis, after substituting $u' = r^m \rchi_1(r)$ and requiring analyticity for $\eta'$, suggests that $v' = r^{m-1} \rchi_2(r)$ and $w' = r^{m-1} \rchi_3(r)$. The addition of Eq.~\eqref{eq:c11t} and Eq.~\eqref{eq:c33t} yields,
\begin{equation}
\overline{a}_1(c_{11}'+c_{33}') + \overline{b}_1c_{22}'-2U_rc_{12}' = (2C_{33}r^{-1}-C_{33r}-C_{11r})v'+2(\mi \alpha C_{11} + C_{12}D)u' + 2\mi n C_{33}w'r^{-1},
\label{eq:c11c33t}
\end{equation}
where $\overline{a}_1 = G_3+(C_{11}+C_{33})G/W$, $\overline{b}_1 = (C_{11}+C_{33})G/W$. The Eqs.~\eqref{eq:c22t}, \eqref{eq:c12t} and \eqref{eq:c11c33t}, can be solved and yield
\begin{equation}
c_{11}' + c_{33}' = \frac{(\overline{d}_1G_3+2U_r\overline{d}_3)\overline{b}_2 - (\overline{b}_1G_3+2U_r\overline{b}_3)\overline{d}_2}{(\overline{a}_1G_3+2U_rC_{12}G/W)\overline{b}_2 - (\overline{b}_1G_3+2U_r\overline{b}_3)C_{22}G/W},
\label{eq:c11c33}
\end{equation}
where $\overline{b}_2 = G_3+C_{22}G/W$, $\overline{b}_3 = C_{12}G/W - U_r$ and $\overline{d}_1$ is the 
RHS of Eq.~\eqref{eq:c11c33t}. The functions $\overline{d}_2$ and $\overline{d}_3$ are the terms on the RHS of Eq.~\eqref{eq:c22t} and Eq.~\eqref{eq:c12t}, respectively, that have explicit dependence on velocity fluctuations. Substituting the variables $u'$, $v'$ and $w'$ in terms of $\rchi_j(r)$\textsc{\char13}s into Eq.~\eqref{eq:c11c33} and carrying out a Taylor analysis round $r=0$, we find that 
\begin{equation}
c_{11}'+c_{33}' \sim r^{m-2}. \label{eq:c11c33_scale}
\end{equation}
Substituting Eq.~\eqref{eq:c11c33_scale} into Eq.~\eqref{eq:c22t}, we find $c_{22}' \sim r^{m-2}$. Therefore, let $c_{22}' = r^{m-2}\rchi_5(r)$. However, using Eq.~\eqref{eq:c11c33_scale} and $c_{22}' \sim r^{m-2}$ in Eq.~\eqref{eq:c11t} and Eq.~\eqref{eq:c33t} does not predict the leading-order behaviours of $c_{11}'$ and $c_{33}'$. The three possibilities arising from Eq.~\eqref{eq:c11c33_scale} are
\begin{align} 
(c_{11}', c_{33}') =& \left \{
\begin{array}{ll}
(r^{m-2}\rchi_4, r^{m-2}\rchi_6) &\hspace{1.7cm}\mbox{(option 1)},\\
(r^{m-2}\rchi_4, r^{s_1}\rchi_6) &\hspace{1.7cm}\mbox{(option 2)},\\
(r^{s_2}\rchi_4, r^{m-2}\rchi_6) &\hspace{1.7cm}\mbox{(option 3)},
\end{array}
\right . \label{eq:options}
\end{align} 
with $s_1 > m-2$ and $s_2> m-2$. For now, let us assume that option (1) is true. However, we will prove in Sec.~\ref{appsec:parity_solnform}, that the option (3) is in fact the correct one, and that $s_2 = m$.  Using option (1) of Eq.~\eqref{eq:options}, analyses of  Eq.~\eqref{eq:c12t}, Eq.~\eqref{eq:c13t} and Eq.~\eqref{eq:c23t} suggest $c_{12}' = r^{m-1}\rchi_7(r)$, $c_{13}' = r^{m-1}\rchi_8(r)$ and $c_{23}' = r^{m-2}\rchi_9(r)$, respectively.

Let us make the following simplifying substitutions for the mean-flow. Since we are interested only in finding the leading-order behaviour and parities of the unknowns without any intention to know exactly the values of constants $\rchi_{j,k}$ in the Taylor expansion, we can represent the mean-flow variables by the first term in their expansion in powers of $r$ since this term would contain the leading-order behaviour and the parities of them (i.e, the mean-flow variables). In what follows, $H_i$, $i = 1, \cdots, 14$ are real constants. Without any loss of generality for the stated purpose to retain only the leading-order behaviour and parity, we carry out the following assignments, $U(r) := H_1$, $U_r(r):=H_2r$, $F(r):= H_3$, $F_r(r) := H_4r$, $G(r) := H_5$, $G_1(r) := H_6$, $C_{12} := H_7r$, $G_2 := H_8r$, $C_{22} = C_{33} := H_9$, $C_{11} := H_{10}$, $G_3 := H_{11}$, $\mi (\alpha H_1 - \omega) + \alpha^2\beta/Re := H_{12}$, $C_{11r} := H_{13}r$ and $C_{22r} := C_{33r}:= H_{14}r$. (Some of these constants are: $H_2 = -2, H_3 = H_5 = H_9 = H_{10} =1, H_4 = 16W^2L^{-2}; H_7 = -2W; H_{13} = 16 W^2[1-L^{-2}]; H_{14} = -H_4 $).

Now, let us substitute the above forms for unknowns $\boldsymbol{u'}$ and $\textbf{\textit{\textsf{c'}}}$ represented using $\rchi_j(r)$'s into Eqs.~\eqref{eq:umom}--\eqref{eq:wmom} to 
determine the integers $m$ and $s$, and the parities $\rchi_j(r)$, if any.  Then, Eq.~\eqref{eq:umom} becomes
\begin{multline}
\sum_{k=0}^\infty \nolimits \left \{ H_{12} \rchi_{1,k}r^{k+m} -\beta Re^{-1} [(k+m)^2-n^2] \rchi_{1,k}r^{k+m-2} \right \} = \sum_{k=0}^\infty \nolimits \left \{ -H_2\rchi_{2,k}r^{k+m} \right .\\ \left . -\mi \alpha \rchi_{0,k}r^{k+s} + \Gamma\left ( \left [ (H_6+\mi \alpha H_3)\rchi_{4,k} +H_6 \left (\rchi_{5,k} + \rchi_{6,k} \right ) + H_5H_7(k+m-2)\left ( \rchi_{4,k} \right . \right . \right . \right . \\ \left . \left . \left . \left . +\rchi_{5,k}+\rchi_{6,k} \right )  + H_3\left ((k+m)\rchi_{7,k} + \mi n \rchi_{8,k} \right )  \right ] r^{k+m-2} + H_4\rchi_{7,k}r^{k+m} \right ) \right \}.
\label{eq:umom_chi_k}
\end{multline}
In Eq.~\eqref{eq:umom_chi_k}, if $s < m-2$, the equations formed by the coefficients of $r^{k+s}$ form a trivial solution for $\rchi_{0,k}$ for each $k = 0,\cdots, (m-3-s)$. Therefore, without any loss of generality, we can choose, $s = m-2$. If in reality $s>m-2$ while we had chosen $s = m-2$, then this fact will show up by giving the solution for the coefficients $\rchi_{0,k} = 0$ for $k = 0, \cdots, s-m+1$ upon which we will be able to correct this exponent $s$ such that $\rchi_{0,0} \neq 0$. (In fact this happens in the following, and we will see that $s = m$.) To summarize, the solution forms of the unknowns currently stand as
\begin{align}
(p',u', v', w') =& (r^{m-2}\rchi_0, r^{m}\rchi_1,r^{m-1}\rchi_2,r^{m-1}\rchi_3),\label{eq:soln_form_internediate_1} \\
(c_{11}',c_{22}',c_{33}', c_{12}',c_{13}',c_{23}') =& (r^{m-2}\rchi_4, r^{m-2}\rchi_5, r^{m-2}\rchi_6,r^{m-1}\rchi_7,r^{m-1}\rchi_8, r^{m-2}\rchi_9), \label{eq:soln_form_internediate_2}
\end{align}
which will further be refined later. Substituting the above into Eqs.~\eqref{eq:vmom}--\eqref{eq:wmom}, we get
\begin{multline}
\sum_{k=0}^\infty \nolimits \left \{ H_{12} \rchi_{2,k}r^{k+m+1} -\beta Re^{-1}\left  [ [(k+m-1)^2-n^2-1] \rchi_{2,k} - 2\mi n \rchi_{3,k} \right ] r^{k+m-1} \right \} \\= \sum_{k=0}^\infty \nolimits \left \{ -(k+m-2) \rchi_{0,k}r^{k+m-1} + \Gamma\left ( \left [ H_8\left (\rchi_{4,k}+\rchi_{6,k}\right ) + \mi \alpha H_3 \rchi_{7,k} \right . \right . \right . \\ \left . \left . \left .  + (H_8+H_4)\rchi_{5,k}   \right ] r^{k+m+1} + \left [ H_3\left (\rchi_{5,k}-\rchi_{6,k}+\mi n \rchi_{9,k} \right ) \right . \right . \right . \\ \left . \left . \left . + (k+m-2)\left ( (H_5H_9+H_3)\rchi_{5,k} + H_5H_9 \left ( \rchi_{4,k} + \rchi_{6,k} \right ) \right ) \right ] r^{k+m-1} \right ) \right \} \ \text{and}
\label{eq:vmom_chi_k}
\end{multline}
\begin{multline}
\sum_{k=0}^\infty \nolimits \left \{ H_{12} \rchi_{3,k}r^{k+m+1} -\beta Re^{-1}\left  [ [(k+m-1)^2-n^2-1] \rchi_{3,k} + 2\mi n \rchi_{2,k} \right ] r^{k+m-1} \right \} \\ =\sum_{k=0}^\infty \nolimits \left \{ -\mi n \rchi_{0,k}r^{k+m-1} + \Gamma\left ( \left [ H_4\rchi_{9,k}+\mi \alpha H_3 \rchi_{8,k} \right ] r^{k+m+1} + \left [ H_3(k+m)\rchi_{9,k} \right .\right . \right . \\  \left . \left . \left . + \mi n\left ( (H_5H_9+H_3)\rchi_{6,k} + H_5H_9 \left ( \rchi_{4,k} + \rchi_{5,k} \right ) \right ) \right ] r^{k+m-1} \right ) \right \}.
\label{eq:wmom_chi_k}
\end{multline}
Similarly, upon substituting Eq.~\eqref{eq:soln_form_internediate_1} and Eq.~\eqref{eq:soln_form_internediate_2} into Eqs.~\eqref{eq:c11t}--\eqref{eq:c23t}, we get
\begin{align}
A_1\sum_{k=0}^\infty \nolimits \rchi_{4,k} r^{k+m-2} =& \sum_{k=0}^\infty \nolimits \left \{ \left [ -H_{13}\rchi_{2,k} + 2(\mi \alpha H_{10}+H_7(k+m))\rchi_{1,k} \right ]r^{k+m} 
\right . \nonumber \\ & \left .
- H_{10}H_5W^{-1}\left ( \rchi_{5,k} + \rchi_{6,k} \right ) r^{k+m-2} \right \}\\
A_2\sum_{k=0}^\infty \nolimits \rchi_{5,k} r^{k+m-2} =& \sum_{k=0}^\infty \nolimits \left \{  (2\mi\alpha H_7-H_{14})\rchi_{2,k}r^{k+m} + 2H_9(k+m-1)\rchi_{2,k} r^{k+m-2} 
\right . \nonumber \\ & \left .
- H_{9}H_5W^{-1}\left ( \rchi_{4,k} + \rchi_{6,k} \right ) r^{k+m-2} \right \}\\
A_2\sum_{k=0}^\infty \nolimits \rchi_{6,k} r^{k+m-2} =& \sum_{k=0}^\infty \nolimits \left \{  2H_9\left [\rchi_{2,k} +\mi n \rchi_{3,k} \right ]r^{k+m-2}  - H_{14}\rchi_{2,k} r^{k+m} 
\right . \nonumber \\ & \left .
- H_{9}H_5W^{-1}\left ( \rchi_{4,k} + \rchi_{5,k} \right ) r^{k+m-2} \right \}\\
H_{11}\sum_{k=0}^\infty \nolimits \rchi_{7,k} r^{k+m-1} =& \sum_{k=0}^\infty \nolimits \left \{ \left [ H_9(k+m)\rchi_{1,k} + \left (\mi \alpha H_{10} + (k+m-2)H_7\right) \rchi_{2,k} 
\right . \right . \nonumber \\ & \left . \left .
+ \left (H_2-H_7H_5W^{-1} \right )\rchi_{4,k} -H_7H_5W^{-1}\left (\rchi_{4,k} + \rchi_{6,k} \right ) \right ]r^{k+m-1} 
\right . \nonumber \\ & \left .
+\mi\alpha H_7\rchi_{1,k} r^{k+m+1} \right \}\\
H_{11}\sum_{k=0}^\infty \nolimits \rchi_{8,k} r^{k+m-1} =&  \sum_{k=0}^\infty \nolimits \left \{ \left [ \mi n H_9\rchi_{1,k} + \left (\mi \alpha H_{10} + (k+m-2)H_7\right) \rchi_{3,k} 
\right . \right . \nonumber \\ & \left . \left .
+ H_2\rchi_{9,k} \right ] r^{k+m-1} \right \}\\
H_{11}\sum_{k=0}^\infty \nolimits \rchi_{9,k} r^{k+m-2} =& \sum_{k=0}^\infty \nolimits \left \{ H_9\left [ (k+m-2)\rchi_{3,k} + \mi n \rchi_{2,k} \right ]r^{k+m-2} 
\right . \nonumber \\ & \left .
+\mi\alpha H_7\rchi_{3,k} r^{k+m} \right \}, \label{eq:x9k_expanded}
\end{align}
where $A_1 = H_{11}+H_{10}H_{5}W^{-1}$, $A_2 = H_{11}+H_{9}H_{5}W^{-1}$. The continuity equation reads
\begin{equation}
\sum_{k=0}^\infty \nolimits \left \{ \mi\alpha \rchi_{1,k} r^{k+m} + \left [ (m+k)\rchi_{2,k} + \mi n \rchi_{3,k} \right ]r^{m-2+k}  \right \} = 0. \label{eq:cont_chi_k}
\end{equation} 
\subsection{Determination of $m$\label{appsec:detm}}
The leading-order terms of Eq.~\eqref{eq:cont_chi_k} satisfy $m\rchi_{2,0}+\mi n \rchi_{3,0} = 0$. With this information, the leading-order terms of Eq.~\eqref{eq:umom_chi_k} and Eqs.~\eqref{eq:vmom_chi_k}--\eqref{eq:x9k_expanded} read
\begin{align}
-(\beta/Re)(m^2-n^2&)\rchi_{1,0} = -\mi \alpha \rchi_{0,0} + \Gamma\left [  (H_6+\mi \alpha H_3)\rchi_{4,0} +H_6 \left (\rchi_{5,0} + \rchi_{6,0} \right ) \right . \nonumber \\ 
&\left .+ H_5H_7(m-2)\left ( \rchi_{4,0} +\rchi_{5,0}+\rchi_{6,0} \right )  + H_3\left (m\rchi_{7,0} + \mi n \rchi_{8,0} \right )  \right ], \label{eq:umom_chi_0}\\
-(\beta/Re)(m^2-n^2&)\rchi_{2,0} = (2-m) \rchi_{0,0} + \Gamma\left [  H_3 \left (\rchi_{5,0} - \rchi_{6,0} + \mi n \rchi_{9,0} \right ) \right . \nonumber \\ 
&\left .+ H_5H_9(m-2)\left ( \rchi_{4,0} +\rchi_{5,0}+\rchi_{6,0} \right )  + H_3(m-2) \rchi_{5,0}  \right ] \ \text{and} \label{eq:vmom_chi_0}\\
-\mi \beta (mRe)^{-1}(m^2-&n^2)(m-2)\rchi_{2,0} = -\mi n \rchi_{0,0} + \Gamma\left [  H_3 \left ( m \rchi_{9,0} + \mi n \rchi_{6,0} \right )\right . \nonumber \\ 
&\left .+ \mi n H_5H_9\left ( \rchi_{4,0} +\rchi_{5,0}+\rchi_{6,0} \right )    \right ], \label{eq:wmom_chi_0}
\end{align}
\begin{align}
A_1\rchi_{4,0} =& -H_{10}H_5W^{-1}(\rchi_{5,0}+\rchi_{6,0}), \label{eq:x40}\\
A_2\rchi_{5,0} =& 2(m-1)H_9\rchi_{2,0}-H_9H_5W^{-1}(\rchi_{4,0}+\rchi_{6,0}), \label{eq:x50}\\
A_2\rchi_{6,0} =& -2(m-1)H_9\rchi_{2,0}-H_9H_5W^{-1}(\rchi_{4,0}+\rchi_{5,0}), \label{eq:x60}\\
H_{11}\rchi_{7,0}=& mH_9\rchi_{1,0} + [\mi \alpha H_{10} + (m-2)H_7]\rchi_{2,0}+H_2\rchi_{5,0}\nonumber \\ 
& -H_7H_5W^{-1}(\rchi_{4,0}+\rchi_{5,0}+\rchi_{6,0}),\label{eq:x70}\\
H_{11}\rchi_{8,0} =& \mi n H_9\rchi_{1,0} + \mi m n^{-1}[\mi \alpha H_{10}+(m-2)H_7]\rchi_{2,0} + H_2\rchi_{9,0} \ \ \ \text{and}\ \label{eq:x80}\\
H_{11}\rchi_{9,0} =& \mi H_9 n^{-1}(m^2+n^2-2m)\rchi_{2,0}.\label{eq:x90}
\end{align}
This system of Eqs.~\eqref{eq:umom_chi_0}--\eqref{eq:x90} has nine unknowns: $\{ \rchi_{0\text{--}2,0}, \rchi_{4\text{--}9,0} \}$. Therefore, the condition that the determinant has to vanish gives the value for the parameter $m$. However, this $9\times 9$ system can be simplified to a $3\times 3$ system as follows. Adding Eq.~\eqref{eq:x50} and Eq.~\eqref{eq:x60}, we get
\begin{equation}
(A_2 + H_9H_5W^{-1}) (\rchi_{5,0} + \rchi_{6,0} ) = -2H_9H_5W^{-1}\rchi_{4,0}. \label{eq:x50x60}
\end{equation}
Solving Eq.~\eqref{eq:x50x60} and Eq.~\eqref{eq:x40} gives $\rchi_{5,0}+\rchi_{6,0} = 0$ and
\begin{equation} 
\rchi_{4,0} = 0. \label{eq:x40_0}
\end{equation}
Further, $m\times$ Eq.~\eqref{eq:x50} $+\mi n \times $ Eq.~\eqref{eq:x90} results in $m\rchi_{5,0}+\mi n \rchi_{9,0} = (m^2-n^2)H_9H_{11}^{-1}\rchi_{2,0}$. Similarly, $m\times$ Eq.~\eqref{eq:x70} $+\mi n \times $ Eq.~\eqref{eq:x80} results in $m\rchi_{7,0}+\mi n \rchi_{8,0} = (m^2-n^2)[H_9H_{11}^{-1}\rchi_{1,0} + H_2H_9H_{11}^{-2}\rchi_{2,0}]$. Similarly, $m\times$ Eq.~\eqref{eq:x90} $+\mi n \times $ Eq.~\eqref{eq:x60} results in $m\rchi_{9,0}+\mi n \rchi_{6,0} = \mi(m^2-n^2)(m-2)H_9(nH_{11})^{-1}\rchi_{2,0}$. Substituting this information into Eqs.~\eqref{eq:umom_chi_0}--\eqref{eq:wmom_chi_0}, we get 
\begin{align}
-A_3(m^2-n^2)\rchi_{1,0} -A_4(m^2-n^2)\rchi_{2,0} =& -\mi \alpha \rchi_{0,0}, \label{eq:xmom_chi0_simplified}\\
-A_3(m^2-n^2)\rchi_{2,0} =& (2-m) \rchi_{0,0}, \label{eq:ymom_chi0_simplified}\\
A_5(m^2-n^2)(m-2)\rchi_{2,0} =& \rchi_{0,0} \label{eq:zmom_chi0_simplified},
\end{align}
respectively, where $A_3 = \beta Re^{-1} + \Gamma H_3H_9H_{11}^{-1}$, $A_4 = \Gamma H_3H_9H_2H_{11}^{-2}$ and $A_5 = \beta (mRe)^{-1} + \Gamma H_3H_9(nH_{11})^{-1}$. The Eqs.~\eqref{eq:xmom_chi0_simplified}--\eqref{eq:zmom_chi0_simplified} are three homogeneous equations for three unknowns. Therefore, the condition that the determinant has to vanish for non-trivial solution for $\{ \rchi_{0\text{--}2,0}\}$ gives rise to $m = |n|$. However, the current choice of $s=m-2$ needs correction, since Eqs.~\eqref{eq:xmom_chi0_simplified}--\eqref{eq:zmom_chi0_simplified} state that $\rchi_{0,0} = 0$ for this choice of $m=|n|$. This will be addressed later after establishing the parity.

\subsection{Parity and solution form \label{appsec:parity_solnform}}
To determine the parity, if any, of $\rchi_j(r)$'s with respect to $r$, let us look at the equations governing $\rchi_{j,1}$. If these equations would result in trivial solutions, then there would be an even symmetry for $\rchi_j(r)$ as explained below. The next higher order terms of Eq.~\eqref{eq:umom_chi_k} and Eqs.~\eqref{eq:vmom_chi_k}--\eqref{eq:cont_chi_k} forms a linear homogeneous system $L_{ij} \rchi_{j,1} = 0$ where $j = 0\text{--}9$ and $L_{ij}$'s are components of a $10 \times 10$ matrix. Since there are no free undetermined parameters in $L$, its determinant may not vanish. This gives $\rchi_{j,1} = 0$ for each $j$. Since the 10 equations, Eq.~\eqref{eq:umom_chi_k} and Eqs.~\eqref{eq:vmom_chi_k}--\eqref{eq:cont_chi_k} couples only alternate terms in the series expansion, we get the following recursion relations for the Taylor coefficients:
\begin{equation}
\rchi_{i,k+2} = L_{ij}^{(k)}\rchi_{j,k}\ \text{for} \ k = 0,1,\cdots, \label{eq:recur}
\end{equation}
where $L_{ij}^{(k)}$ is the corresponding linear system that could be defined from the Eq.~\eqref{eq:umom_chi_k} and Eqs.~\eqref{eq:vmom_chi_k}--\eqref{eq:cont_chi_k} for every higher order. Equation~\eqref{eq:recur} implies that $\rchi_{j,k} = 0$ for every odd $k$ since $\rchi_{j,1} = 0$ and $\rchi_{j,k}$ are non-trivial for every even $k$ since the solution set, $\{\rchi_{j,0}\}$ is non-trivial. This proves the even parity of $\rchi_j(r)$. 

The leading-order behaviour for $c_{11}'\equiv r^{|n|-2}\rchi_4(r)$ can be corrected as follows. By Eq.~\eqref{eq:x40_0} and the established parity, we have $\rchi_{4,0} = \rchi_{4,1} = 0$. This suggests that $\rchi_4(r)$ itself is having a leading-order behaviour of $r^2$ which can used to redefine $c_{11}'$ as $c_{11}'\equiv r^{|n|}\rchi_4(r)$. Note that such redefinition does not affect the solution obtained for $m(=|n|)$. Such a redefinition implies that the LHS of Eq.~\eqref{eq:x40} is zero at the lowest order, suggesting $\rchi_{5,0}+\rchi_{6,0} = 0$, which in turn is a result that is consistent with Eq.~\eqref{eq:x50x60}. 

Further, $\rchi_{0,0} = \rchi_{0,1} = 0$, since the indicial solution $m =|n|$ renders $\rchi_{0,0} = 0$ by any of the Eqs.~\eqref{eq:xmom_chi0_simplified}--\eqref{eq:zmom_chi0_simplified}. This implies that we can adjust the exponents of $r$ in the definition given be Eqs.~\eqref{eq:soln_form_internediate_1}--\eqref{eq:soln_form_internediate_2} by increasing them by 2 for $p'$ without any loss of generality.  Such an increase in the exponent for $r$ from $m-2$ to $m$ would not affect the very system given by Eqs.~\eqref{eq:xmom_chi0_simplified}--\eqref{eq:zmom_chi0_simplified} that determined the value of $m$ as $m=|n|$, except that the RHS of this system would now be zeros at the lowest order, which still determines $m = |n|$. The final solution form is 
\begin{align}
(p',u', v', w') =& (r^{|n|}\rchi_0, r^{|n|}\rchi_1,r^{|n|-1}\rchi_2,r^{|n|-1}\rchi_3) \ \text{and}  \label{eq:soln_form_1} \\
(c_{11}',c_{22}',c_{33}', c_{12}',c_{13}',c_{23}') =& (r^{|n|}\rchi_4, r^{|n|-2}\rchi_5, r^{|n|-2}\rchi_6,r^{|n|-1}\rchi_7,r^{|n|-1}\rchi_8, r^{|n|-2}\rchi_9), \label{eq:soln_form_2}
\end{align}
with Taylor expansions for $\rchi_j(r)$ as $\rchi_j(r) = \sum_{k =0}^\infty \rchi_{j,2k}r^{2k}$. 

\section{The case of $n =\pm 1$}\label{appB}
For the ease of proceeding further, note that $H_2 = -2$, $H_7 = -2W$ and $H_3= H_5 = H_9 = H_{10} = 1$.

Since the governing equations for this case are not different from those we saw for the case of $|n|\geq2$, the analysis we carried out for that case still applies here, except that a modification is needed for the exponents of the leading order behaviour of $c_{22}', c_{23}'$ and $c_{33}'$. These components of $\textbf{\textit{\textsf{c'}}}$ acquires new set of exponents for this case of $|n|=1$, which are in fact an increment by two to the exponents predicted by substituting $|n|=1$ into Eq.~\ref{eq:soln_form_2}. This substitution give for these variables the following leading order behaviour.
\begin{equation}
(c_{22}',c_{33}', c_{23}') = (r^{-1}\rchi_5, r^{-1}\rchi_6,r^{-1}\rchi_9). \label{eq:soln_form_modn_1_incorrect}
\end{equation}
The above Eq.~(\ref{eq:soln_form_modn_1_incorrect}) apparently predicts that $c_{22},c_{33}'$ and $c_{23}'$ are singular at the origin. However as we will show below the constant term in the Taylor expansion of $\rchi_5$, $\rchi_6$ and $\rchi_9$ around $r=0$ are zero. 

Substituting $|n|=1$ into Eqs.~\ref{eq:soln_form_1}-\ref{eq:soln_form_2}, the variables in $(u', v',w', c_{11}')$ take the form,
\begin{equation}
(u', v',w', c_{11}') = (r\rchi_1, \rchi_2,\rchi_3,r\rchi_4). \label{eq:soln_form_modn_1_1}
\end{equation}
These variables are regular. 

Since we found in Appendix A, that $m = |n|$, let us substitute $m = 1$ in Eq.~(\ref{eq:x90}) we get $\rchi_{9,0}=0$. By the even parity nature, we get $\rchi_{9,1} =0$. This suggest that we can increase the exponent for leading order behaviour of $c_{23}'$ by 2 in Eq.~(\ref{eq:soln_form_modn_1_incorrect}). Therefore, the correct behaviour is $c_{23}' =r\rchi_9$. Now we focus our attention on $c_{22}'$ and $c_{33}'$. Adding Eq.~(\ref{eq:x50}) and Eq.~(\ref{eq:x60}), and substituting for $\rchi_{4,0}$ from Eq.~(\ref{eq:x40}) we get $\rchi_{5,0}+\rchi_{6,0}=0$.
By substituting this result into Eq.~(\ref{eq:x50}) and Eq.~(\ref{eq:x60}), and noting that $m=1$, we get $\rchi_{5,0}=\rchi_{6,0}=0$. Again, by the even parity of $\rchi_j(r)$, $\rchi_{5,1}=\rchi_{6,1}=0$. this allow us to increase the leading order exponents of $r$ in $c_{22}'$ and $c_{33}'$ by two. Therefore we arrive at the relations $c_{22}' =r\rchi_5$ and $c_{33}' =r\rchi_6$.

\section{Case $n = 0$ and $\alpha \neq 0$}\label{appC}
The governing equations after substituting the mean flow variables by their respective leading order terms become
\begin{align}
\mi\alpha u' =& -Dv' - r^{-1}v' \label{eq:cont}\\
\mathcal{L}_1u'=&-H_2rv'-\mi\alpha p' +\Gamma\left [ (H_6 +\mi\alpha H_3)c_{11}' + H_6(c_{22}' + c_{33}')  \right . \nonumber \\
&\left . + H_7H_5rD(c_{11}'+c_{22}'+c_{33}')  + (H_4r + H_3r^{-1} + H_3D)c_{12}' \right ] \label{eq:umom_n0al_neq_0}\\
\mathcal{L}_1v'=&-p'_r - \beta(r^2\Rey)^{-1}v' +\Gamma\left [ H_8rc_{11}' + (H_8r+H_4r+H_3r^{-1})c_{22}' + \mi\alpha H_3c_{12}'  \right . \nonumber \\
&\left .  + (H_8r-H_3r^{-1})c_{33}' + H_9H_5D(c_{11}'+c_{33}') +(H_3+H_9H_5)Dc_{22}'  \right ]
\label{eq:vmom_n0al_neq_0}\\
\mathcal{L}_1w'=&- \beta(r^2\Rey)^{-1}w' +\Gamma\left [\mi \alpha H_3 c_{13}' +(H_4r + 2H_3r^{-1}+H_3 D)c_{23}' \right ]\label{eq:wmom_n0al_neq_0}\\
H_{11}c_{11}' =& -H_{13}rv' + 2(\mi \alpha H_{10}u' + H_7ru_r' + H_2r c_{12}') -H_{10}H_5W^{-1}(c_{11}'\!+\! c_{22}'\!+ \!c_{33}')\label{eq:c11t_n0al_neq_0}\\
H_{11}c_{12}' =& (\mi \alpha H_{10}-H_7+ H_7rD)v' + (\mi\alpha H_7r+ H_9D)u' + (H_2 - H_7H_5W^{-1})rc_{22}'  \nonumber \\
&   -H_7H_5W^{-1}r(c_{11}'+ c_{33}')\label{eq:c12t_n0al_neq_0}\\
H_{11}c_{13}' =& (\mi \alpha H_{10}-H_7+ H_7rD)w' + H_2rc_{23}'  \label{eq:c13t_n0al_neq_0}\\
H_{11}c_{22}' =& ( 2\mi \alpha H_7r -H_{14}r + 2H_9D )v' -H_9H_5W^{-1}(c_{11}'+ c_{22}'+ c_{33}')\label{eq:c22t_n0al_neq_0}\\
H_{11}c_{23}' =& (\mi \alpha H_7r-H_9r^{-1}+ H_9D)w'  \label{eq:c23t_n0al_neq_0}\\
H_{11}c_{33}' =& (2H_9r^{-1}-H_{14}r)v' -H_9H_5W^{-1}(c_{11}'+ c_{22}'+ c_{33}')\label{eq:c33t_n0al_neq_0}
\end{align}
where, the differential operator, $\mathcal{L}_1 = \mi(\alpha H_1 -\omega ) - \beta\Rey^{-1}\left [ D^2 + 
r^{-1}D -\alpha^2 \right ]$. The Eqs.~(\ref{eq:cont})-(\ref{eq:c33t_n0al_neq_0}) is a system for 10 unknowns namely, $p',u',v',w',c_{11}', c_{22}', c_{33}', c_{12}', c_{13}'$, and $c_{23}'$. It should be noted that this system forms two different decoupled system for two sets of solutions, namely, $\boldsymbol{\rchi^{(1)}}\equiv (p',u',v',c_{11}',c_{22}',c_{33}',c_{12}')$ and $\boldsymbol{\rchi^{(2)}}\equiv (w',c_{13}',c_{23}')$. The $\boldsymbol{\rchi^{(1)}}$ is governed by Eqs.~(\ref{eq:cont})-(\ref{eq:vmom_n0al_neq_0}), Eqs.~(\ref{eq:c11t_n0al_neq_0})-(\ref{eq:c12t_n0al_neq_0}), Eq.~(\ref{eq:c22t_n0al_neq_0}) and Eq.~(\ref{eq:c33t_n0al_neq_0}), while the set $\boldsymbol{\rchi^{(2)}}$ is governed by Eq.~(\ref{eq:wmom_n0al_neq_0}), Eq.~(\ref{eq:c13t_n0al_neq_0}) and Eq.~(\ref{eq:c23t_n0al_neq_0}). 

\subsection{The parity and behaviour of $\boldsymbol{\rchi^{(1)}}$}

Let $p'=r^s\rchi_0(r)$ and $u'=r^m\rchi_1(r)$, where the exponents, $m$ and $s$ are to be determined. From Eq.~(\ref{eq:cont}), we get $v' = r^{m+1}\rchi_2(r)$, by demanding analyticity of $v'$ at $r=0$.  Now we proceed by using the values of some of the constants $H_j$'s listed in Appendix~\ref{appB}. 

Adding Eq.~(\ref{eq:c11t_n0al_neq_0}) and Eq.~(\ref{eq:c33t_n0al_neq_0}) we get,
\begin{align}
(H_{11} +2W^{-1})(c_{11}'+c_{33}') + 2W^{-1}c_{22}' + 4rc_{12}' =&  2r^{m} (\mi \alpha -2Wm -2WrD)\rchi_1\nonumber \\
& + r^m[2-r^2(H_{13}+H_{14})]\rchi_2. \label{eq:c11c33_temp}
\end{align}
The Eq.~(\ref{eq:c12t_n0al_neq_0}) and Eq.~(\ref{eq:c22t_n0al_neq_0}) can be written as
\begin{align}
-2 r(c_{11}'+ c_{33}') + H_{11}c_{12}' =& r^{m+1}(\mi \alpha -2Wm  -2W rD)\rchi_2\nonumber \\
&  + r^{m-1}(-2W\mi\alpha r^2+ m+rD)\rchi_1, \label{eq:c12t_temp}\\
W^{-1}(c_{11}'+ c_{33}') + (H_{11}+W^{-1})c_{22}' =& r^m[ -(4\mi \alpha W+H_{14})r^2 + 2(m+1)+ 2rD ]\rchi_2. \label{eq:c22t_temp}
\end{align}
The three Eqs.~(\ref{eq:c11c33_temp})-(\ref{eq:c22t_temp}) can be solved for three unknown $c_{11}'+c_{33}'$, $c_{22}'$ and $c_{12}'$, and can be found to the leading order that $c_{11}'+c_{33}'\sim r^m$, $c_{22}'\sim r^m$ and $c_{12}'\sim r^{m-1}$ by analysis. Substituting these behaviours in Eq.~(\ref{eq:c11t_n0al_neq_0}) and Eq.~(\ref{eq:c33t_n0al_neq_0}), we find that $c_{11}'\sim r^m$ and $c_{33}'\sim r^m$, respectively. Therefore, let $c_{11}' = r^m\rchi_4(r)$, $c_{22}' = r^m\rchi_5(r)$, $c_{33}' = r^m\rchi_6(r)$, $c_{12}' = r^{m-1}\rchi_7(r)$. 

Substituting these expressions for $p',u',v',c_{11}',c_{22}',c_{33}'$ and $c_{12}'$ into Eqs.~(\ref{eq:cont})-(\ref{eq:vmom_n0al_neq_0}), Eqs.~(\ref{eq:c11t_n0al_neq_0})-(\ref{eq:c12t_n0al_neq_0}), Eq.~(\ref{eq:c22t_n0al_neq_0}) and Eq.~(\ref{eq:c33t_n0al_neq_0}), and representing $\rchi_j(r) = \sum_{k=0}^\infty r^k\rchi_{j,k}$, we get,
\begin{equation}
\mi\alpha \rchi_{1,k} = - (m+k+2)\rchi_{2,k}\ \ \  \text{for each} \ k, \label{eq:cont_expanded}\\
\end{equation}
\begin{multline}
\sum_{k = 0}^\infty \nolimits r^{k+m} \left \{ H_{12} - \beta\Rey^{-1}(m+k)^2r^{-2} \right \}\rchi_{1,k}=\sum_{k = 0}^\infty \nolimits \left \{ 2r^{m+k+2}\rchi_{2,k}-\mi\alpha r^{s+k}\rchi_{0,k} \right . \\ \left . +\Gamma r^{m+k}\left [ (H_6 +\mi\alpha )\rchi_{4,k} + H_6(\rchi_{5,k} + \rchi_{6,k})  + (m+k)(\rchi_{4,k}+\rchi_{5,k}+\rchi_{6,k}) 
\right . \right . \\ \left . \left . 
+ (H_4 + [m+k]r^{-2})\rchi_{7,k} \right ] \right \}, \label{eq:umom_expanded}
\end{multline}
\begin{multline}
\sum_{k = 0}^\infty \nolimits r^{k+m+1}\! \left \{ H_{12}\! - \!\beta\Rey^{-1}r^{-2}\left [ (m\!+\!k\!+\!1)^2 \!-\! 1\right ] \right \}\rchi_{2,k}= \sum_{k = 0}^\infty \nolimits \! \left \{ -(s\!+\!k)r^{s+k-1}\rchi_{0,k}
\right . \\ \left .
+\Gamma r^{m+k+1}\left [ H_8\rchi_{4,k} + (H_8+H_4+r^{-2})\rchi_{5,k} + \mi\alpha r^{-2}\rchi_{7,k}  + (H_8-r^{-2})\rchi_{6,k} 
\right . \right . \\ \left . \left . 
+ (m+k)r^{-2}(\rchi_{4,k}+\rchi_{6,k} +2\rchi_{5,k})  \right ] \right \},
\label{eq:vmom_expanded}
\end{multline}
\begin{align}
H_{11}\sum_{k = 0}^\infty \nolimits r^{k+m} \rchi_{4,k} =& \sum_{k = 0}^\infty \nolimits r^{k+m} \left \{ -H_{13}r^2\rchi_{2,k} + 2[\mi \alpha + H_7(m+k)]\rchi_{1,k} -4\rchi_{7,k}
\right . \nonumber \\ 
&\left . -W^{-1}(\rchi_{4,k}\!+\! \rchi_{5,k}\!+ \!\rchi_{6,k}) \right \}, \label{eq:c11t_expanded}\\
H_{11}\sum_{k = 0}^\infty \nolimits r^{k+m-1} \rchi_{7,k} =& \sum_{k = 0}^\infty \nolimits r^{k+m-1} \left \{ [\mi \alpha+ 2W - 2W(m+k+1)]r^2\rchi_{2,k} 
\right . \nonumber \\ & \left . 
+ (-2\mi\alpha Wr^2+ m+k)\rchi_{1,k}  +2r^2(\rchi_{4,k}+ \rchi_{6,k}) \right \},\label{eq:c12t_expanded}\\
H_{11}\sum_{k = 0}^\infty \nolimits r^{k+m} \rchi_{5,k} =& \sum_{k = 0}^\infty \nolimits r^{k+m} \left \{ [-(4\mi \alpha W +H_{14})r^2 + 2(m+k+1) ]\rchi_{2,k} 
\right . \nonumber \\ & \left . 
-W^{-1}(\rchi_{4,k}+ \rchi_{5,k}+ \rchi_{6,k}) \right \}, \label{eq:c22t_expanded}\\
H_{11}\sum_{k = 0}^\infty \nolimits r^{k+m} \rchi_{6,k} =& \sum_{k = 0}^\infty \nolimits r^{k+m} \left \{ (2-H_{14}r^2)\rchi_{2,k} -W^{-1}(\rchi_{4,k}+ \rchi_{5,k}+ \rchi_{6,k}) \right \} \label{eq:c33t_expanded}
\end{align}
In Eq.~(\ref{eq:vmom_expanded}), the lowest order non-pressure terms are of $O(r^{m-1})$ that occurs for $k = 0$. This suggests that the exponent $s\nless m$, since that would result in $\rchi_{0,k} = 0$ for $k \leq m-s-1$. This shows that $s\geq m$. Let us assume that $s = m$ for now. (If the actual value of $s$ is such that $s > m$, this will show up with the result $\rchi_{0,k} = 0$ for $k \leq s-m-1$, which can be used to correct the value of $s$.) After substituting, $\rchi_{2,0}=-\mi\alpha (m+2)^{-1}\rchi_{1,0}$ from Eq.~(\ref{eq:cont_expanded}) into Eqs.~(\ref{eq:umom_expanded})-(\ref{eq:c33t_expanded}), the leading order terms form the system,
\begin{align}
-\beta\Rey^{-1}m^2\rchi_{1,0}=& \Gamma m\rchi_{7,0}, \label{eq:umom_lead}\\
\mi\alpha\beta[(m+2)\Rey]^{-1}\left [ (m\!+\!1)^2 \!-\!1\right ] \rchi_{1,0} =&  -m\rchi_{0,0}
+\Gamma\left [ \rchi_{5,0}\! +\! \mi\alpha\rchi_{7,0}\!  -\!\rchi_{6,0}
\right . \nonumber \\  & \left .
\!+\! m(\rchi_{4,0}\!+\!\rchi_{6,0} \!+\!2\rchi_{5,0})  \right ] ,
\label{eq:vmom_lead}
\end{align}
\begin{align}
H_{11} \rchi_{4,0} =& 2(\mi \alpha \rchi_{1,0} -2Wm\rchi_{1,0} -2\rchi_{7,0}) -W^{-1}(\rchi_{4,0}\!+\! \rchi_{5,0}\!+ \!\rchi_{6,0}) , \label{eq:c11t_lead}\\
H_{11}\rchi_{7,0} =& m\rchi_{1,0} ,\label{eq:c12t_lead}\\
H_{11}\rchi_{5,0} =&-2\mi\alpha (m+2)^{-1}(m+1) \rchi_{1,0}  
-W^{-1}(\rchi_{4,0}+ \rchi_{5,0}+ \rchi_{6,0}), \label{eq:c22t_lead}\\
H_{11}\rchi_{6,0} =&-2\mi\alpha (m+2)^{-1}\rchi_{1,0}-W^{-1}(\rchi_{4,0}+ \rchi_{5,0}+ \rchi_{6,0})\label{eq:c33t_lead}
\end{align}
A set of non-trivial solution for the system given by Eqs.~(\ref{eq:umom_lead})-(\ref{eq:c33t_lead}) would exist when $m$ takes value such that the determinant of the coefficient matrix would vanish, which turns out to be $m = 0$. Note that for this value of $m$, the Eq.~(\ref{eq:vmom_lead}) allow any value for $\rchi_{0,0}$, that is $\rchi_{0,0}$ is non-trivial. This confirms that the choice $s=m(=0)$ is the exponent for correct leading order behaviour for pressure around the pipe centre.

Since the set of Eqs.~(\ref{eq:umom_expanded})-(\ref{eq:c33t_expanded}) couples every alternate higher order equations in powers of $r$, the next higher order equations for $\rchi_{j,1}$ do not depend on $\rchi_{j,0}$. Since the free parameter $m$ has been already fixed to obtain non-trivial solution for $\rchi_{j,0}$, the linear system $\rchi_{j,1}$, $L_{ij}\rchi_{j,1} = 0$ that governs $\rchi_{j,1}$ would allow only trivial values, i.e., $\rchi_{j,1}=0$ for each $j \in \{0,1,2,4\text{--}7\}$.

Let's pay a particular attention to $\rchi_7(r)$. Since $\rchi_{7,0}=0$ by Eq.~(\ref{eq:c12t_lead}) and $\rchi_{7,1}=0$ by parity, we redefine the earlier assumed exponent for the leading order behaviour for $c_{12}'$. Earlier we had chosen it as $c_{12}'=r^{m-1}\rchi_7(r)$, which can be modified as $c_{12}'=r\rchi_7(r)$. Note that such definition does not affect the value determined for m. Such redefinition would imply that the LHS of Eq.~(\ref{eq:c12t_lead}) would be zero in the lowest order, which gives $m=0$, a result that is consistent with earlier finding. Therefore, we arrive at
\begin{equation}
(p',u', v',c_{11}',c_{22}',c_{33}', c_{12}') = (\rchi_0, \rchi_1,r\rchi_2,\rchi_4, \rchi_5, \rchi_6,r\rchi_7). \label{eq:soln_form_n0alneq0_part1}
\end{equation}

\subsection{The leading order behaviour and parities of $\boldsymbol{\rchi^{(2)}}$}
Let $w' = r^q\rchi_3(r)$ where $q$ is a positive integer required by analyticity that will be determined later. The Eq.~(\ref{eq:c23t_n0al_neq_0}) and Eq.~(\ref{eq:c13t_n0al_neq_0}) suggest the leading order behaviours for $c_{23}'$ and $c_{13}'$ as $c_{23}' = r^{q-1}\rchi_9(r)$ and $c_{13}' = r^q\rchi_8(r)$, respectively. Upon substituting these expressions in Eq.~(\ref{eq:wmom_n0al_neq_0}), Eq.~(\ref{eq:c13t_n0al_neq_0}) and Eq.~(\ref{eq:c23t_n0al_neq_0}),and representing $\rchi_j(r)$'s by their respective Taylor expansion around $r=0$, we get,
\begin{multline}
\sum_{k = 0}^\infty \nolimits r^{k+q} \left \{H_{12}\rchi_{3,k} - \beta\Rey^{-1} r^{-2}\left [ (q+k)^2-1\right ] \rchi_{3,k} \right \}
\\
= \Gamma \sum_{k = 0}^\infty \nolimits r^{k+q} \left \{ \mi \alpha\rchi_{8,k}+H_4\rchi_{9,k} +r^{-2}(k+q+1)\rchi_{9,k}  \right \},\label{eq:wmom_expanded}
\end{multline}
\begin{align}
H_{11}\sum_{k = 0}^\infty \nolimits r^{k+q}\rchi_{8,k} =& \sum_{k = 0}^\infty \nolimits r^{k+q} \left \{[\mi \alpha-2W(k+q-1)]\rchi_{3,k} -2\rchi_{9,k} \right \},  \label{eq:c13t_expanded}\\
H_{11}\sum_{k = 0}^\infty \nolimits r^{k+q-1}\rchi_{9,k} =& \sum_{k = 0}^\infty \nolimits r^{k+q-1} \left \{[-2\mi \alpha Wr^2-(k+q-1)]\rchi_{3,k} \right \}.  \label{eq:c23t_expanded}
\end{align}
The leading order terms form the system,
\begin{align}
- \beta\Rey^{-1} ( q^2-1) \rchi_{3,0} =& \Gamma  (q+1)\rchi_{9,0} ,\label{eq:wmom_lead}\\
H_{11}  \rchi_{8,0} =&   [\mi \alpha-2W(q-1)]\rchi_{3,0} -2\rchi_{9,0},  \label{eq:c13t_lead}\\
H_{11}\rchi_{9,0} =& -(q-1)\rchi_{3,0}.  \label{eq:c23t_lead}
\end{align}
The determinant of coefficient matrix of the system given by Eq.~(\ref{eq:wmom_lead}) and Eq.~(\ref{eq:c13t_lead}) reads that $q = 1$. Similar to earlier cases, we have $\rchi_{3,1}=\rchi_{8,1}=\rchi_{9,1}=0$. 

Now, since $\rchi_{9,0}=0$ Eq.~(\ref{eq:c23t_lead}), and $\rchi_{9,1}=0$ by parity, we can redefine the exponent of the leading order behaviour for $c_{23}'$ by increasing it by two. Therefore, we have,
\begin{equation}
(w',c_{13}', c_{23}') = (r\rchi_3, r\rchi_8,r^2\rchi_9). \label{eq:soln_form_n0alneq0_part2}
\end{equation}
Note that such redefinition of $c_{23}'$ as $c_{23}' = r^2\rchi_9$ would not affect the determined value for $q(=1)$. Such redefinition would mean the LHS of Eq.~(\ref{eq:c23t_lead}) would be zero at the lowest order, which again gives $q=1$. 

The even-party of each of $\rchi_j(r)$ is also immediate by the same arguments used for the case of $|n|\geq 2$ and mentioned in the main paper.

\section{Case $n = 0$ and $\alpha = 0$}\label{appD}
For this case, the continuity reads, $rDv'+v' = 0$ which has $v'=0$ as the solution that is analytic at $r=0$. The other variables, follow the same leading order behaviour as in the previous case of $n = 0$ and $\alpha \neq 0$ except for $c_{11}', c_{22}', c_{33}'$ and $c_{13}'$. In the previous case, we found that $c_{11}'=r^m\rchi_4(r), c_{22}'=r^m\rchi_4(r), c_{33}'=r^m\rchi_4(r)$ with $m= 0$. With this value for $m$ for the present case of $n = 0$ and $\alpha = 0$, the Eq.~(\ref{eq:c11t_lead}), Eq.~(\ref{eq:c22t_lead}) and Eq.~(\ref{eq:c33t_lead}) become
\begin{align}
H_{11} \rchi_{4,0} =& -W^{-1}(\rchi_{4,0}\!+\! \rchi_{5,0}\!+ \!\rchi_{6,0}) , \label{eq:c11tal0_lead}\\
H_{11}\rchi_{5,0} =&-W^{-1}(\rchi_{4,0}+ \rchi_{5,0}+ \rchi_{6,0}), \label{eq:c22tal0_lead}\\
H_{11}\rchi_{6,0} =&-W^{-1}(\rchi_{4,0}+ \rchi_{5,0}+ \rchi_{6,0}),\label{eq:c33tal0_lead}
\end{align}
which has the solution, $\rchi_{4,0} =\rchi_{5,0} =\rchi_{6,0}=0$. Due to the established even parity, $\rchi_{4,1} =\rchi_{5,1} =\rchi_{6,1}=0$. Therefore the correct leading order behaviour for this case is $c_{11}'=r^2\rchi_4(r), c_{22}'=r^2\rchi_4(r)$ and $c_{33}'=r^2\rchi_4(r)$. Similarly, the leading order exponent for $c_{13}'$ needs to incremented by two since the Eq.~(\ref{eq:c13t_lead}) reads that $\rchi_{8,0} = 0$ for $\alpha =0$. Therefore, $c_{13}'=r^3\rchi_8(r)$.

\section{Proof using the results of Fourier analysis for $n=0$ \label{appE}}

For $n=0$, the Fourier analysis by Lewis and Bellan~\cite{lewis1990physical} predicts $v' = r\rchi_2$ and $w' = r\rchi_3$ where $\rchi_2$ and $\rchi_3$ are even functions of $r$. Since these results are independent of physical problems, these are applicable in the current non-Newtonian flow as well. 

Substituting these results into the continuity equation, $\mi\alpha u' = -Dv' - r^{-1}v'$, we get $u' = \rchi_1$, which is also an even function of $r$. Let the Taylor expansions for $\rchi_j(r)$ (j = 1, 2, 3) be  $\rchi_j(r) = \sum_{k =0}^\infty \rchi_{j,2k}r^{2k}$.

Substituting these expressions for $u'$, $v'$ and $w'$ into the Eqs.~(\ref{eq:c11t_n0al_neq_0})-(\ref{eq:c33t_n0al_neq_0}), and noting that $H_2 = -2$, $H_7 = -2W$and $H_3= H_5 = H_9 = H_{10} = 1$, and retaining only the leading order terms, we get

\begin{align}
H_{11}c_{11}' =& -K_1r^2 + \alpha K_2 -4r c_{12}' -W^{-1}(c_{11}'\!+\! c_{22}'\!+ \!c_{33}'),\label{eq:c11t_n0al_neq_0_fourier}\\
H_{11}c_{12}' =& K_3r +  2r(c_{11}'+ c_{33}'),\label{eq:c12t_n0al_neq_0_fourier}\\
H_{11}c_{13}' =& \alpha r K_4 -2rc_{23}',  \label{eq:c13t_n0al_neq_0_fourier}\\
H_{11}c_{22}' =&  K_5r^2 + K_7  -W^{-1}(c_{11}'+ c_{22}'+ c_{33}'),\label{eq:c22t_n0al_neq_0_fourier}\\
H_{11}c_{23}' =& K_6r^2, \label{eq:c23t_n0al_neq_0_fourier}\\
H_{11}c_{33}' =& K_7 + K_8r^2 -W^{-1}(c_{11}'+ c_{22}'+ c_{33}'),\label{eq:c33t_n0al_neq_0_fourier}
\end{align}
where the constants, $K_i$'s are given by $K_1 = (H_{13}\rchi_{2,0} + 2H_{7}\rchi_{1,2})$, $K_2 = 2\mi \rchi_{1,0}$, $K_3 = \mi \alpha \rchi_{2,0} + \mi\alpha H_7\rchi_{1,0}+ 2\rchi_{1,2}$, $K_4 = \mi \rchi_{3,0}$, $K_5 = 2\mi \alpha H_7\rchi_{2,0} -H_{14}\rchi_{2,0}$, $k_6 = (\mi \alpha H_7\rchi_{3,0} + 2\rchi_{3,2})$, $K_7 = 2\rchi_{2,0}$, $K_8 = -H_{14}\rchi_{2,0}$. The linear system given by Eqs.~(\ref{eq:c11t_n0al_neq_0_fourier})-(\ref{eq:c33t_n0al_neq_0_fourier}) for $c_{ij}'$ can be solved to show that
\begin{equation} 
(c_{11}',c_{22}',c_{33}', c_{12}',c_{13}',c_{23}') = (\rchi_4, \rchi_5, \rchi_6, r\rchi_7, r\rchi_8, r^2\rchi_9),
\end{equation}
when $\alpha \neq 0$. Though $c_{11}'$, $c_{22}'$ and $c_{33}'$ are non-zero, the trace $c_{ii}'$ vanishes at the centreline. The sum of Eq.~\ref{eq:c11t_n0al_neq_0_fourier}, Eq.~\ref{eq:c22t_n0al_neq_0_fourier} and Eq.~\ref{eq:c33t_n0al_neq_0_fourier} shows that $c_{ii}'\sim \mi\alpha \rchi_{1,0}+2\rchi_{2,0}$ to the leading order. However, $\mi\alpha \rchi_{1,0}+2\rchi_{2,0} = 0$ by the virtue for the continuity equation to the leading order. 

When $\alpha = 0$, the Eq.~(\ref{eq:c11t_n0al_neq_0_fourier})-(\ref{eq:c33t_n0al_neq_0_fourier}) become,
\begin{align}
H_{11}c_{11}' =& -K_1r^2 -4r c_{12}' -W^{-1}(c_{11}'\!+\! c_{22}'\!+ \!c_{33}'),\label{eq:c11t_n0al_0_fourier}\\
H_{11}c_{12}' =& K_3r +  2r(c_{11}'+ c_{33}'),\label{eq:c12t_n0al_0_fourier}\\
H_{11}c_{13}' =& -2rc_{23}',  \label{eq:c13t_n0al_0_fourier}\\
H_{11}c_{22}' =& -W^{-1}(c_{11}'+ c_{22}'+ c_{33}'),\label{eq:c22t_n0al_0_fourier}\\
H_{11}c_{23}' =& K_6r^2, \label{eq:c23t_n0al_0_fourier}\\
H_{11}c_{33}' =& -W^{-1}(c_{11}'+ c_{22}'+ c_{33}'),\label{eq:c33t_n0al_0_fourier}
\end{align}
where the information $v' = 0$, i.e., $\rchi_{2,j} = 0$ from continuity equation is employed. The similar analysis of Eq.~(\ref{eq:c11t_n0al_0_fourier})-(\ref{eq:c33t_n0al_0_fourier}) reveals
\begin{equation} 
(c_{11}',c_{22}',c_{33}', c_{12}',c_{13}',c_{23}') = (r^2\rchi_4, r^2\rchi_5, r^2\rchi_6, r\rchi_7, r^3\rchi_8, r^2\rchi_9).
\end{equation}
Similar analysis can be performed for the cases, $|n|=1$ and $|n|\geq2$ using the predictions for $v'$ and $w'$ from Fourier analysis of \cite{lewis1990physical}.
}
\bibliography{polymer_ansatz}
\end{document}